\documentclass[12pt]{iopart}
\def\newblock{\hskip .11em plus .33em minus .07em}

\expandafter\let\csname equation*\endcsname\relax
\expandafter\let\csname endequation*\endcsname\relax
\usepackage{amsmath}
\usepackage{amssymb}
\usepackage{amsxtra,color}
\usepackage{latexsym}
\usepackage{graphicx}
\usepackage{natbib}

\newcommand{\bra}[1]{\langle  \: #1 \: |}
\newcommand{\ket}[1]{| \: #1 \: \rangle}
\newcommand{\braket}[2]{\langle \: #1 \: | \: #2 \: \rangle}
\newcommand{\braOket}[3]{\langle \: #1 \: | \: #2 \:| \: #3 \: \rangle}

\newcommand{\braOketRed}[3]{\langle \: #1 \: || \: #2 \: || \: #3 \: \rangle}

\newcommand{\bracorrket}[2]{\langle \: #1 \: |  r^{-1}_{12}   | \: #2 \: \rangle}

\newcommand{\intsum}[1]{\sum_{#1} \! \! \! \! \! \! \! \! \int }

\newcommand{\threej}[6]{\left( \begin{matrix}
				#1 & #2 & #3 \\
				#4 & #5 & #6 \\
\end{matrix} \right)}

\begin{document}

%\author{J.~M.~Dahlstr\"om$^{1,2,3}$ and E.~Lindroth$^{1}$ \\
%\small $^1$Department of Physics, Stockholm University, \\ 
%\small AlbaNova University Center, SE-106 91 Stockholm, Sweden}

%\date{}

\maketitle

\title{Study of attosecond delays using \\ % in laser-assisted photoionization  \\ 
 perturbation diagrams and exterior complex scaling}

\author{J.~M.~Dahlstr\"om$^{1,2,3}$ and E.~Lindroth$^{1}$}
\address{$^1$Department of Physics, Stockholm University, AlbaNova University Center, SE-106 91 Stockholm, Sweden}
\address{$^2$Max Planck Institute for the Physics of Complex Systems, Noethnitzerstr. 38, 01187 Dresden} 
\address{$^3$Center for Free-Electron Laser Science, Luruper Chaussee 149, 22761 Hamburg}
\ead{marcus.dahlstrom@fysik.su.se}

%\pacs{32.80.Rm, 32.80.Qk, 42.65.Ky}

\begin{abstract}

‌%\input{abstracttext}

We describe in detail how attosecond delays in laser-assisted photoionization can be computed using  perturbation theory based on two-photon matrix elements. Special emphasis is laid on above-threshold ionization, where the electron interacts with an infrared field after photoionization by an extreme ultraviolet field. Correlation effects are introduced using diagrammatic many-body theory to the level of the random-phase approximation with exchange (RPAE). Our aim is to provide an {\it ab initio} route to correlated multi-photon processes that are required for an accurate description of experiments on the attosecond time scale. Here, our results are focused on photoionization of the $M$-shell of argon atoms, where experiments have been carried out using the so-called RABITT technique. An influence of autoionizing resonances in attosecond delay measurements is observed. Further, it is shown that the delay depends on both detection angle of the photoelectron and energy of the probe photon.

\end{abstract}

\maketitle 

% INTRODUCTION 
\section{Introduction}
In the light of modern experiments on the attosecond timescale 
\cite{AgostiniRPP2004,KlingARPC2008,KrauszRMP2009}
it is interesting to revisit and to adapt 
some of the most well-established  theoretical approximations developed 
during the golden age of many-body atomic physics in the 70's and 80's 
\cite{mbpt,Amusia1990}. 
Attosecond experiments are carried out using a wide range of radiation parameters 
from photoionization by attosecond extreme ultraviolet (XUV) pulses (weak-field high-frequency regime) \cite{PaulScience2001} 
to high-order harmonic generation (HHG)  (strong-field low-frequency regime) \cite{LewensteinPRA1994}. 
Methods to evaluate and probe the HHG process remains a busy subject on its own, 
in particular for gaining structural information of the targets in laser-driven recollisions 
\cite{HaesslerJPB2011,ShafirNature2012}. 
Multielectron effects are weak compared to the interaction with lasers in the strong-field limit. This has allowed for the application of the strong field approximation (SFA) in  numerous scenarios including intense lasers, c.f. the tutorial by Madsen \cite{MadsenAJP2005}. 
One of these branches is the `soft-photon approximation' for ionization 
of atoms by an extreme-ultraviolet (XUV) attosecond pulse 
in the presence of a synchronized (assisting) infra-red (IR) laser-field \cite{MaquetJMO2007,AlvaroNJP2013}. 
In the above mentioned works, however, the use of SFA implies that the interactions between the photoelectron and the ion are neglected. 
This includes both attosecond delays from the Coulomb electron motion and correlation effects. 
The influence of the asymptotic Coulomb potential, known as the laser-Coulomb coupling, has been studied within the Eikonal-Volkov Approximation (EVA) \cite{SmirnovaPRA2008,ZhangPRA2010,IvanovPRL2012}, 
but also by full numerical simulations, c.f. Ref.~\cite{NagelePRA2011,PazourekFD2013}.
Experiments using laser-assisted photoionization by attosecond pulses have evidenced that such atomic effects are measurable as relative `atomic  delays' observed from different initial orbitals of the atoms 
\cite{Schultze2010a,KlunderPRL2011,GuenotPRA2012}. 
It has been shown that the delays, $\tau_\mathrm{A}$, 
can be separated as the sum of the Wigner-like delay, $\tau_\mathrm{W}$, 
of the electron from the XUV photoionization process 
plus a universal contribution from the interaction with the IR field, called the continuum--continuum delay (CC) or Coulomb-laser coupling delay, $\tau_\mathrm{cc}$
\cite{KlunderPRL2011,NagelePRA2011,DahlstromCP2013} 
\begin{equation}
\tau_\mathrm{A}=\tau_\mathrm{W}+\tau_\mathrm{cc}.
\label{delays}
\end{equation} 
In our earlier work we demonstrated the validity of Eq.~(\ref{delays}) 
for electrons in unstructured continuum from noble gas atoms \cite{DahlstromPRA2012}. 
The same equation has been shown to hold on helium with some modifications to account for shake-up processes \cite{PazourekPRL2012}. 
However, slow delay structures in the continuum, 
which arise at autoionizing resonances \cite{fanoPR1961}, 
are expected to break the validity of  Eq.~(\ref{delays}) 
\cite{CarettePRA2013}.   

The goal of this paper is to present a computer program, based on atomic many-body theory \cite{mbpt}, 
that has allowed us to compute atomic delays including electron--electron correlation 
and the ionic potential for noble gas atoms \cite{DahlstromPRA2012}.   
The first photon is absorbed from the attosecond XUV field (pump) while the second photon is exchanged with an IR field (probe). 
The atomic delays can then be computed by adding different interfering quantum paths leading to the same final state using the Reconstruction of Attosecond Beating by Two-photon Interference Transitions (RABITT) method, c.f.~Ref.~\cite{MullerAPB2002,TomaJPB2002}. 
While correlation effects can be included to infinite order in terms of Coulomb interactions, the interactions with the fields must be restricted to lowest-order perturbation theory in our approach. 
These limitations are not very restrictive for attosecond metrology based on {\it above}-threshold ionization 
\cite{DahlstromJPB2012}, 
and certainly not for the RABITT method, where both pump and probe are perturbative fields.  
%Provided that the intensity of the probe field is not too strong, 
Similarly,
the atomic response from the attosecond streak-camera method \cite{ItataniPRL2002,MairessePRA2005}
can be quantitatively calculated using two-photon matrix-elements \cite{DahlstromCP2013}.  
Recently, this close connection between the RABITT and the streak-camera method has been demonstrated using full numerical propagation of photoelectron wavepackets \cite{PazourekFD2013}. 
The Phase Retrieval by Omega Oscillation Filtering (PROOF) method, which was originally formulated using the SFA 
\cite{ChiniOE2010}, has been improved to include atomic effects using two-photon matrix elements (IPROOF) \cite{LaurentOE2013}. 
As the IPROOF method is not limited to symmetric pulse trains it may serve as a quantitative tool 
for studying non-symmetric pulse trains 
\cite{MauritssonPRL2006,DahlstromJPB2011,LaurentPRL2012}. 
Common to all above mentioned approaches is that the photoelectron is assumed to transfer directly into the continuum after absorption of the XUV photon. This  allows for the use of Fermi's golden rule. It is, in principle, necessary to convolute the two-photon matrix elements over XUV and IR bandwidth, but such a procedure can be avoided if the IR field is assumed to be monochromatic \cite{DahlstromJPB2012}. 
In contrast, phase effects from the field envelopes in {\it below}-threshold two-photon ionization or on resonances in the continuum have been identified, c.f.~Ref.~\cite{SwobodaPRL2010,CaillatPRL2011}. Theoretical studies by Ishikawa and Ueda show that ultra-short pulse envelopes (from few femtosecond to attosecond duration) gives rise to phase effects that affect also the photoelectron angular distribution \cite{IshikawaPRL2012,IshikawaAS2013}. In order to understand such detailed effects either full time-dependent simulations must be performed, or, all field-convolutions in perturbation theory must be performed. While such convolutions are beyond the scope of the present work, we want to stress they could, in principle, be performed by repeated evaluation and summation of two-photon matrix elements over the incident field frequencies. 

% First XUV-XUV HHG  KobayashiOL1998
%               FEL  MoshammerOE2011

It is clear that non-linear optics using attosecond XUV fields have opened up for new ways to probe photoionization dynamics, but let us start with a detour in Sec.~\ref{sec_crossar} and present a short review of traditional cross-section measurements in argon, as a means of verifying our theoretical method with well established experimental data and the implications of various theoretical approximations. 
In Sec.~\ref{sec_method} we then describe in full length the implementation for two-photon above-threshold ionization matrix elements. In Sec.~\ref{sec_numres} we present results for the corresponding attosecond delays in laser-assisted photoionization including 
angle-resolved detection (Sec.~\ref{sec_detang}), 
autoionizing resonances (Sec.~\ref{sec_fano}) and 
the different probe wavelength (Sec.~\ref{sec_probewave}). 
Finally, in Sec.~\ref{sec_conc}, we present our conclusion.

\subsection{Review of M-shell photoelectron cross-sections in argon}
\label{sec_crossar}
\begin{figure}[h]
	\centering
	\includegraphics[width=0.45\textwidth]{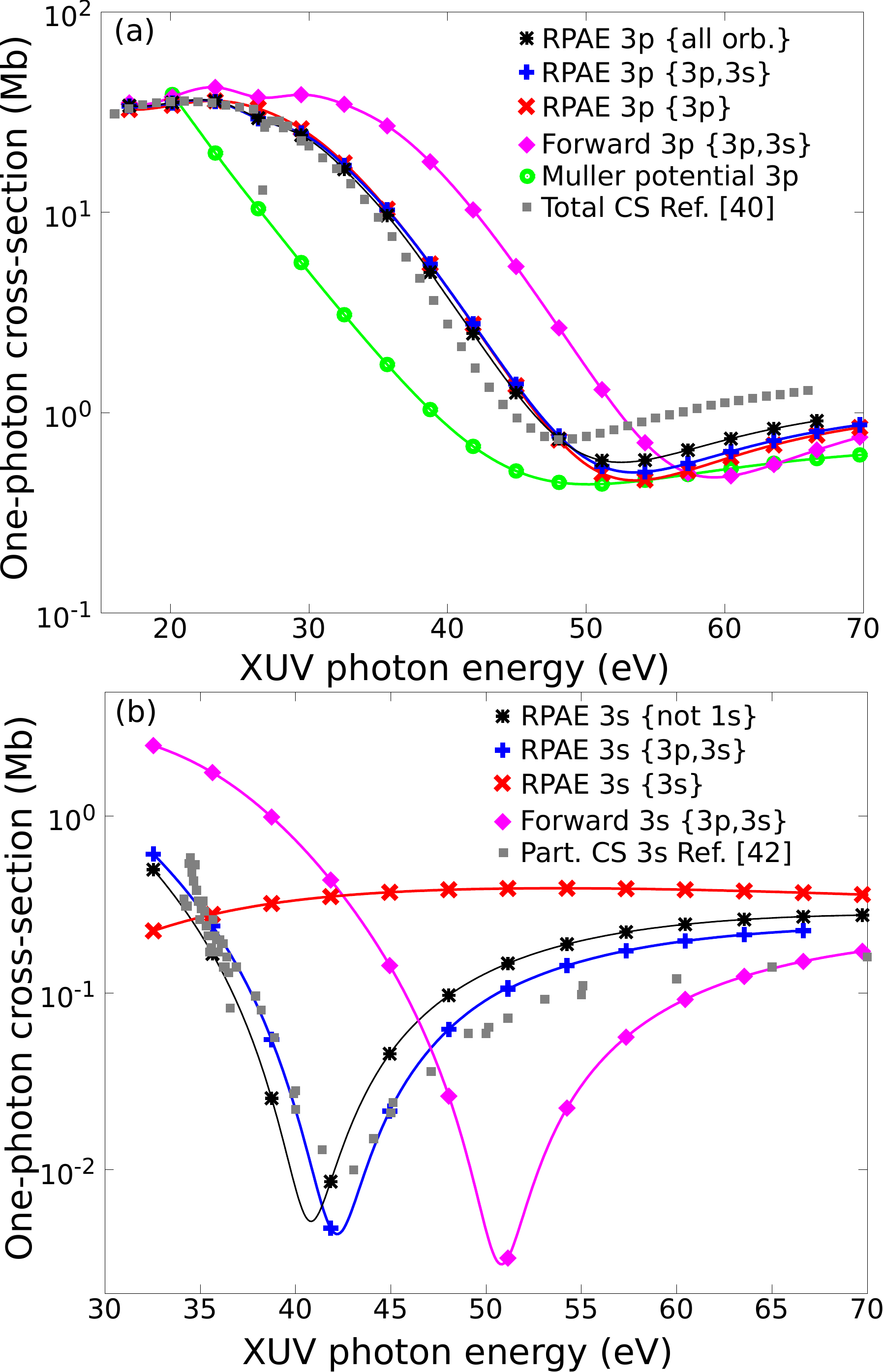}
	\caption{
(Color online)
(a) Theoretical partial photoionization cross-section for Ar$3p^{-1}$ compared to the total argon photoionization cross-section measurements from Ref.~\cite{SamsonJES2002}.
(b) Theoretical partial photoionization cross-section for Ar$3s^{-1}$ compared to partial cross-section measurements from 
Ref.~\cite{MobusPRA1993}.
The curly brackets list correlated orbitals within the RPAE model.
} 	
\label{fig_cross} 
\end{figure}

Photoionization experiments of noble gas atoms by synchrotron light has provided essential benchmark data for the development of atomic many-body theory over the past decades. The total cross-sections of noble gas atoms have been determined with a high accuracy of a few percent in the  threshold region using XUV radiation \cite{SamsonJES2002}. Close to the ionization threshold, the total photoionization cross-section will be equal to the partial cross-section from the most weakly bound orbital. 
In Fig.~\ref{fig_cross}(a) we show that the total experimental cross-section by Samson and Stolte (grey $\opensquare$) is in excellent agreement with the partial cross-section from the $3p$ orbital ($\times,+,*$) calculated with the so-called random-phase approximation with exchange (RPAE):  
(red $\times$) includes correlation only within the $3p$ orbital; 
(blue $+$) includes correlation among $3p$ and $3s$ orbitals; and 
(black $*$) includes correlation between all atomic orbitals. 
The RPAE method is explained in detail by Amusia in Ref.~\cite{Amusia1990}.   
More generally, any theory including ground state correlation effects (in particular the $3p^6$ -- $3p^{4}3d^2$ interaction), will suffice to get good agreement in the close-to-threshold region, as pointed out by Manson in his review on atomic photoelectron spectroscopy \cite{Manson1978II}. 
The so-called Hartree-Fock approximation for photoionization, which is an independent electron theory used by Kennedy and Manson \cite{kennedyPRA1972}, does not include such correlation effects and it is therefore not a good approximation close to the ionization threshold of argon (not shown). 
At higher photon energies more channels open, corresponding to various excited states of the remaining ion. Here, the total cross-section may differ from the first partial cross-section. In the case of argon, photoionization starts at an XUV photon energy of $15.76\,$eV from the $3p$ orbital. Following the $3s$-autoionizing resonances \cite{SorensenPRA1994}, the next channel opens at $29.24\,$eV with partial photoionization from the $3s$ orbital. Above $38.60\,$eV, photoionization satellites are populated with doubly excited $3p$ electrons, 
c.f. the diagrammatic work of Wijesundera and Kelly \cite{kellyPRA1989} and the recent work by 
Carette et al. \cite{CarettePRA2013}. 
The `disagreement' between the positions of the Cooper minimum 
in Fig.~\ref{fig_cross}~(a) for the total cross section ($\opensquare$ at $48.3\,$eV) 
and the partial RPAE calculation ($*$ at $53.5\,$eV)  
is explained by the additional ionization channels contributing to the 
total cross section.
The `exact' position of the Cooper minimum from $3p$ was determined to be $53.8\pm0.7\,$eV by Higuet et al. using the HHG process 
\cite{HiguetPRA2011}. 
Interestingly, this value is expected to differ from both total and partial cross-sections because it corresponds to an 
{\it angle-resolved} partial cross-section, 
which additionally depends on the amplitude of the returning wave packet 
in the recombination step 
of the HHG process \cite{ChenPRA2009}. 
Theoretically, we find that the partial cross-section from the $3p$ orbital with angle-resolved  
photoelectrons restricted to the polarization axis is located at $54.3\,$eV 
(not shown). This is, indeed, in good agreement with the HHG measurement \cite{HiguetPRA2011} 
(no correction for the amplitude of the returning wave packet was made).

% MULLER
In Fig.~\ref{fig_cross}~(a) we show the $3p$ partial cross-section computed using the `Muller potential' (green $\opencircle$) \cite{MullerPRA1999}:
\begin{equation}
V_\mathrm{Muller}(r)=-\frac{1}{r}(1+5.4e^{-r}+11.6e^{-3.682r}),
\label{mullerpot}
\end{equation}
which reduces to the unscreened nuclear potential,
$V_\mathrm{Muller}(r)\approx -18/r$, in the limit of a small radial distance;  
and to a singly charged ion, 
$V_\mathrm{Muller}(r)\approx -1/r$, 
in the limit of a large radial distance. 
 While this single-active electron potential was originally designed to model strong-field physics with low-frequency laser fields, it was soon also implemented in the theoretical analysis of XUV attosecond pulse trains within the so-called RABITT scheme \cite{PaulScience2001,TomaJPB2002}. As expected \cite{HiguetPRA2011}, 
the lack of correlation effects leads to quite poor agreement with the experimental data close to the ionization threshold of the one-photon XUV photoionization process.
% MAURITSSON
It is possible to optimize the pseudo-potential in order to better match the experimental cross-sections, as was done by Mauritsson et al. in Ref.~\cite{MauritssonPRA2005}. 
%Unfortunately, we did not succeed in implementing the Argon potential from this work, because we did not find a suitable analytical representation for the steep potential variations close to the core. 
In macroscopic-scale calculations, the sheer efficiency of pseudo-potentials has made it possible to couple the atomic response 
of high-order harmonic generation to simulations using the Maxwell wave equation \cite{GaardeJPB2008}. For more detailed atomic calculations, methods that include correlation effects consistently are required.

%TDCIS
Recently, an ab initio time-dependent configuration interaction singles (TDCIS) method has been developed to go beyond the single-active electron approximation in describing the interaction of strong fields with atoms \cite{GreenmanPRA2010}. In a recent review article, based on the PhD thesis of Pabst \cite{Pabst2013}, the  TDCIS method was also applied to XUV photoionization cross-sections. By comparison with the RPAE results, we find that the TDCIS calculations were in worse agreement with experimental results. 
In order to understand this discrepancy, we present in Fig~\ref{fig_cross}(a) the result of an RPAE calculation with inter-shell correlation included but ground-state correlation excluded (magenta $\opendiamond$), which is in reasonable agreement with the TDCIS result 
(compare with inter-channel CIS in Fig.~3.4~(a) in Ref.~\cite{Pabst2013}).  
We have verified that both inter-shell and intra-shell RPAE calculations are in good agreement for ionization from the $3p$-orbital, which is in contrast to the TDCIS method that shows marked different results between inter-shell and intra-shell correlation. 
In this way, we suggest that it is the missing double-electron excitations, from the ground-state correlation of the RPAE method, that is the source of discrepancy in the TDCIS method.
 
% Partial 3s cross-section:
While early photoionization experiments were limited to measuring the total cross-section by detecting the total number of ions produced by the incident radiation, partial cross-sections are today accessible using photoelectron time-of-flight spectrometers 
\cite{LynchPLA1973,SamsonPRL1974,MobusPRA1993}.   
As predicted by Amusia, the photoionization from outer $s$ subshells involves strong intershell correlation leading to a dramatic change in the partial cross-section \cite{AmusiaPLA1972}. In Fig.~\ref{fig_cross}~(b) we show this effect by comparison of partial cross-sections including only intra-shell correlation within the $3s$ (red $\times$) with those including inter-shell correlation (blue $+$ and black $*$). A minimum in the partial cross-section is observed at $\sim42\,$eV only when inter-shell correlation is included. The experimental data by M\"obus et al. is best matched by the RPAE model that includes only the $M$-shell electrons ($3p$ and $3s$). This minimum has been interpreted as a destructive interference effect between the paths from the $3p$ orbital and the $3s$ orbital \cite{ChangPRA1978,sahaPRA1989}. 
In a more na\"ive picture, the `$3s$ Cooper minimum' is a copy of the $3p$ Cooper minimum shifted in photon energy by approximately the difference in orbital energies.
The many-body perturbation theory by Chang \cite{ChangPRA1978} and the multi-configuration Hartree-Fock method by Saha \cite{sahaPRA1989} give the best agreement with the experiment close to the Cooper minimum \cite{MobusPRA1993}. Our results including correlation within the $M$-shell (blue $+$) show similar good agreement for the position of the Cooper minimum. At higher energies our calculations show an exaggerated cross-section, which was also observed using the MBPT (in the length gauge) and the MCHF method.
Our implementation of the RPAE method, as well as the recent work on attosecond delays by Kheifets \cite{KheifetsPRA2013}, contains all angular-momentum allowed transitions and is closely related to the MBPT calculations by Chang \cite{ChangPRA1978}.

When ground-state correlation effects are neglected from the $3s$-photoionization process (see Sec.~\ref{sec_diagcorr}), 
we observe a large shift of the entire cross-section to higher energies by roughly $9\,$eV.  Interestingly, when we include also the $L-$shell (black $*$), composed of $2p$ and $2s$ orbitals, we observe a small shift toward lower energies, away from the experimental data. To our knowledge, this inner-core effect has not been pointed out before. We have verified that the addition of K-shell correlation does not further modify the cross-section.

% METHOD !!!

\section{Method}
\label{sec_method}

% ADD THIS STUFF:

In this section we describe in detail how attosecond time delays in photoionization from noble gas atoms can be computed using correlated two-photon (XUV+IR) above-threshold matrix elements \cite{DahlstromPRA2012}. 
In fact, time is not a direct observable in quantum mechanics, so there is no straight-forward way to calculate time delays. 
The work presented here is of more pragmatic nature because we emulate directly how the experimental measurements are made rather than trying to assign it a specific physical interpretation.   
The ``delay in photoionization'' is measured experimentally by recording a spectrogram of photoelectrons as a function kinetic energy (or momentum) and the delay between the XUV field and the IR field. This means that the quantum mechanical observable is the probability for creating a photoelectron with a given energy (or momentum) in the presence of the fields. 
If the incident XUV field is an attosecond pulse train, then it consists of odd high-order harmonics of the IR field:
$\Omega_>=(2N+1)\omega$ and $\Omega_<=(2N-1)\omega$, which can lead to the final sideband state with energy $\epsilon_q=2N\omega-|\epsilon_a|$ by absorption and emission of an additional IR photon, respectively 
(see Fig.~\ref{fig_photonpicture}(a) centre and right blocks).
In this setup, which is called RABITT \cite{MullerAPB2002,TomaJPB2002}, the signal arises from interference between two different two-photon processes. 
If instead the incident XUV field is composed of both even and odd high-order harmonics, or of a continuous XUV spectra, then the so-called (I)PROOF signal \cite{ChiniOE2010,LaurentOE2013} arises from interference between one-photon and two-photon matrix elements (see Fig.~\ref{fig_photonpicture}(a), e.g. left and centre blocks). 
Finally, the streak-camera signal, which can be interpreted as the result of classical motion of the electron in the fields \cite{NagelePRA2011}, involves all such combinations of photon diagrams (including also higher order processes). 
In Refs.~\cite{DahlstromJPB2012,DahlstromCP2013} it is shown that all such photon paths suffer approximately the same phase-shift due to the two-photon matrix element, and so the atomic delay is universal for all methods. As we will show here, the quantitative delays do depend on how the electron detection is made (energy or angle-resolved momentum). 
In addition, all photon-paths much be convoluted over the bandwidth of the XUV and IR fields. This can make the use of photon matrix elements time-consuming for numerical calculations, so we have assumed that the IR field is monochromatic.  The frequency of the XUV photon is then determined, e.g. $\Omega_>=\epsilon_q-\epsilon_a-\omega$ for the absorption path. Note that this trick is valid not only for odd XUV harmonics but also for continuum XUV radiation. Here, in the spirit of the RABITT method, we compute only atomic delays from odd high-order harmonics. 

We begin by a review of the uncorrelated matrix elements (Sec.~\ref{sec_2mat}), which follows mostly the earlier work on hydrogen \cite{DahlstromCP2013}.  
We then provide details of the numerical implementation, which includes  Hartree-Fock for the atomic potential and a complex scaled B-spline basis (Sec.~\ref{sec_numint}).  
Correlation effects are then included to the level of RPAE for screening of the XUV photon (Sec.~\ref{sec_diagcorr}). 
Finally we describe how to compute the continuum--continuum dipole transitions in asymptotic Coulomb potentials using exterior complex scaling (Sec.~\ref{sec_IRinducedtrans}). 

\subsection{Two-photon matrix elements for ionization}
\label{sec_2mat}
\begin{figure}[ht]
	\centering
	\includegraphics[width=\textwidth]{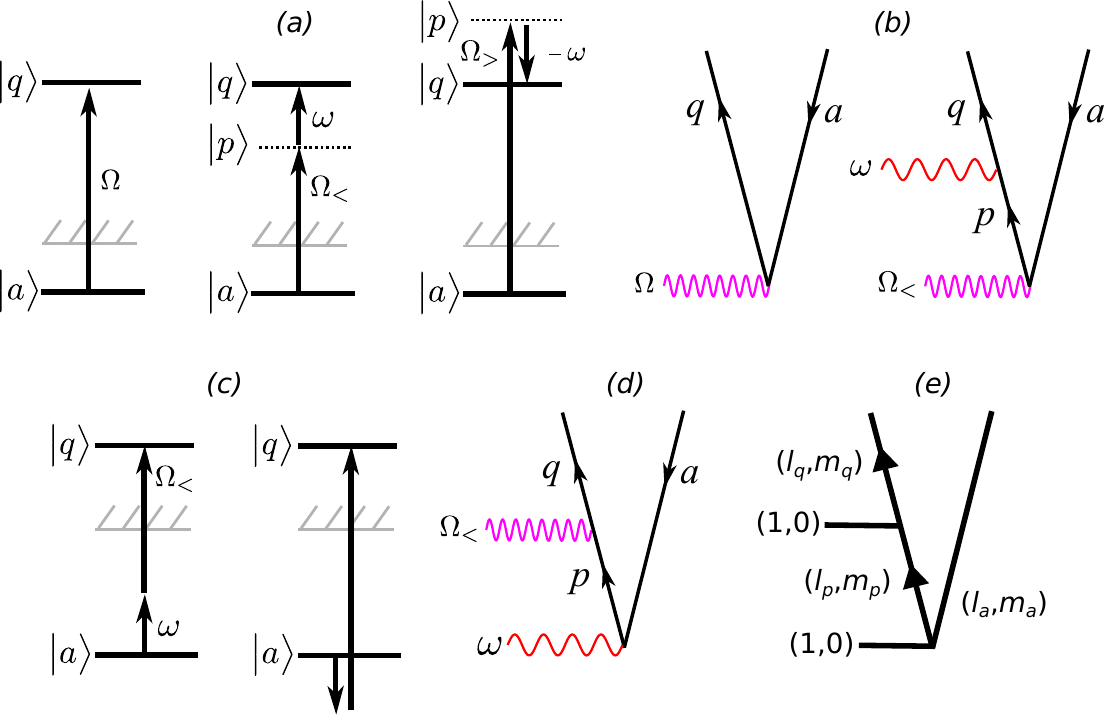}
	\caption{\label{fig_photonpicture} 
(Color online)
Diagrams for lowest-order ionization processes:
(a)--(b) one photon (XUV) and normal time-order (XUV--IR) two-photon process; and 
(c)--(d) reverse time-order (IR--XUV) two-photon process illustrated 
as photon picture and perturbation diagrams, respectively.
(e) Angular momentum diagram for absorption of two photons
with linear polarization, see Ref.~\cite{mbpt} for notation.
} 	
\end{figure}

The matrix element for absorption of two photons (XUV and IR) from the atomic state $a$ to the final state $q$ is 
\cite{TomaJPB2002,DahlstromCP2013}
%(do we drop the $-i$ prefactor?)
\begin{equation}
M(q,\omega,\Omega,a)=
\frac{1}{i}E_\omega E_\Omega
\lim_{\varepsilon\rightarrow 0^+}
\intsum{p}
\frac{ \braOket{q}{z}{p} \braOket{p}{z}{a} }
{\epsilon_a+\Omega-\epsilon_p+i\varepsilon},
\label{MqoOa}
\end{equation}
where the linear polarization 
of the fields is along the $\mathbf{\hat z}$ direction; and $E_\Omega$ and $E_\omega$ are field amplitudes for XUV and IR, respectively.   
Atomic units are used: $\hbar=e=m=1/4\pi\epsilon_0=1$, unless otherwise stated. 
Energy conservation for the two-photon transition  must be satisfied: 
$\epsilon_q-\epsilon_a=\Omega+\omega$, where $\epsilon_i$ are single particle energies and 
where $\Omega$ and $\omega$ correspond to the XUV and IR field, respectively, 
as shown in Fig.~\ref{fig_photonpicture}~(a). 
The matrix element in Eq.~(\ref{MqoOa}) 
can be drawn as the perturbation diagram shown in 
Fig.~\ref{fig_photonpicture}~(b) [right block]. 
Here, the $\Omega$ photon creates an electron-hole pair, 
e.g. an electron wave packet in the excited states $p$ 
and a hole in the atomic state $a$; then the IR photon is absorbed by the electron in a transition to the final state $q$. 
The one-photon analogue is shown for reference in Fig.~\ref{fig_photonpicture}~(b) [left block].

Assume that the initial and final states are pure partial wave states: 
$\braket{\mathbf{r}}{a}=R_{n_a,\ell_a}(r)Y_{\ell_a,m_a}( \mathbf{\hat r})$ 
and 
$\braket{\mathbf{r}}{q}=R_{k_q,\ell_q}(r)Y_{\ell_q,m_q}( \mathbf{\hat r})$ 
with well defined angular momentum and magnetic quantum numbers, 
$\ell_i$ and $m_i$. 
We will sometimes use  the following short-hand notation for the radial states:
$R_{n_i,\ell_i}\equiv R_i$. 
The final state for the photoelectron with energy 
$\epsilon_q=k_q^2/2$  and angular momentum $\ell_q$ in a Coulomb potential 
has the asymptotic form \cite{Friedrich2006}
\begin{equation}
\lim_{r\rightarrow\infty}R_{q}(r)=
\frac{N_q}{r}
\sin\left[k_qr+k_q^{-1}\ln2k_qr-\pi \ell_q/2+\sigma_{q}+\delta_{q}\right],
\label{finalasymp}
\end{equation}
where
$-\pi \ell_q/2$ 
is the centrifugal phase factor,  
$\sigma_q=\arg[\Gamma(\ell_q+1-i/k_q)]$ 
is the Coulomb phase and 
$\delta_q$ 
is the phase-shift due to additional short-range atomic interactions.
The wavefunction is normalized to energy scale: 
$N_q=\sqrt{2/\pi k_q}$.
It is required to include some effective potential that goes beyond
the Hartree-Fock potential
in order to obtain the correct asymptotic form in Eq.~(\ref{finalasymp}), 
c.f. Ref.~\cite{Amusia1990}. 
This is because the excited 
electrons should `feel' an ion with one electron missing, 
as will be reviewed in Sec.~\ref{sec_numint}. 

In length gauge the dipole operator for linear polarization is 
$\hat{\mathbf z} \cdot \mathbf{r}=z= r\cos\theta=rC_0^1$, where $C_0^1$ is the zeroth component of the rank one C-tensor.
Each term in the matrix elements can be expressed using angular momentum theory as 
\begin{equation}
\frac{ \braOket{q}{z}{p} \braOket{p}{z}{a} }
{\epsilon_a+\Omega-\epsilon_p}= 
\frac{ \braOketRed{q}{r}{p} \braOketRed{p}{r}{a}}
{\epsilon_a+\Omega-\epsilon_p}
\times D_0(l_a,l_p,l_q;m_a,m_p,m_q),
\label{twophotonterm}
\end{equation}
where the reduced matrix elements are
\begin{equation}
\braOketRed{i}{r}{j}=
(-1)^{-l_i}
\sqrt{(2l_i+1)(2l_j+1)}
\threej{l_i}{1}{l_j}{0}{0}{0}
\times
\braOket{R_i}{r}{R_j}
\label{reducedmatrix}
\end{equation}
and the factor that depends on the magnetic quantum numbers is
\begin{equation}
D_0=
\sum_{m_p=-l_p}^{l_p}
(-1)^{l_q+l_p-m_q-m_p}
\times
\threej{l_q}{1}{l_p}{-m_q}{0}{m_p}
\threej{l_p}{1}{l_a}{-m_p}{0}{m_a}
,
\label{angulardiagram}
\end{equation}
corresponding to the angular-momentum diagram presented 
in Fig.~\ref{fig_photonpicture}~(e) \cite{mbpt}.
The well-known dipole selection rules follow from 
the properties of the 3j--symbols: 
$\ell_p=\ell_a\pm1\ge0$ and 
$\ell_q=\ell_a,\ell_a\pm2\ge0$; 
while the use of linear polarization imposes that 
the magnetic quantum number of electron and hole must be equal,  
$m_a=m_p=m_q$ 
(the sum in Eq.~(\ref{angulardiagram}) is thus reduced to one term).
We note that the explicit phase factors present in 
Eqs.~(\ref{reducedmatrix}) and (\ref{angulardiagram}) 
will cancel and not contribute to the overall phase 
of the matrix element in Eq.~(\ref{twophotonterm}).

\subsubsection{XUV-initiated electron wave packet}
\label{sec_xuvinit}

Provided that the energy of the XUV photon is sufficient 
to ionize the atom for the initial state, 
$\Omega>|\epsilon_a|$, 
the photoelectrons will be an outgoing wave packet
when it absorbs the IR photon \cite{TomaJPB2002,DahlstromCP2013}. 
By summation (and integration) over all 
terms for excited $p$ states of a given angular momentum $l_{p}$
in Eq.~(\ref{MqoOa}),
the two-photon process is reformulated as a 
one-photon process from an intermediate perturbed wavefunction 
to the final state:
\begin{equation}
\lim_{\varepsilon\rightarrow 0^+}
\intsum{p}
%\sum_{n_{p}}^{\mathrm{exc}}
\frac{ \braOket{q}{z}{p} \braOket{p}{z}{a} }
{\epsilon_a+\Omega-\epsilon_{p}+i\varepsilon}
\equiv 
 \braOketRed{q}{r}{\rho_{\tilde p}}
\times D_0(l_a,l_p,l_q;m_a,m_p,m_q),
\label{twophotonpert}
\end{equation} 
where the index $\tilde p$ on the right hand side is the `on-shell' value
determined by  the pole of Eq.~(\ref{twophotonpert}), 
$\epsilon_{\tilde p}\equiv k_{\tilde p}^2/2=\epsilon_a+\Omega$. 
The perturbed wavefunction takes the asymptotic form of a complex outgoing wave packet  
\cite{DahlstromCP2013}
\begin{equation}
\lim_{r\rightarrow\infty}\rho_{\tilde p}(r)=
-\pi \frac{N_{\tilde p}}{r}
\exp[i(k_{\tilde p}r+k_{\tilde p}^{-1}\ln2k_{\tilde p}r-\pi l_p/2+\sigma_{\tilde p}+\delta_{\tilde p})]
\times 
\braOketRed{\tilde p}{r}{a}.
\label{pertasymp}
\end{equation}
The strength of the one-photon ionization process is determined by the real XUV photon dipole matrix element, 
$\braOketRed{\tilde  p}{r}{a}$. 
By introducing correlation into the ionization process the dipole matrix element may acquire a complex phase 
leading to a phase-shift of the outgoing photoelectron \cite{KheifetsPRL2010,GuenotPRA2012,KheifetsPRA2013}. 
Using our method for including correlation effects, described in Sec.~\ref{sec_diagcorr}, 
we will directly observe this phase-shift of the outgoing photoelectron.

\subsubsection{IR-initiated electron wave packet}
\label{sec_irinit}

The process where the IR photon is absorbed first and the XUV photon
is absorbed last is shown in Fig.~\ref{fig_photonpicture}~(c) and (d)
as photon diagram and perturbation diagram, respectively. 
The IR photon energy is too small to create any ionization on its own, 
$\omega\ll|\epsilon_a|$, so the photoelectron wave packet will be localized predominantly 
close to the atom in the intermediate state.
The corresponding matrix element will 
not have a pole in the continuum,
\begin{equation}
M_{q,\Omega,\omega,a}=
\frac{1}{i}E_\omega E_\Omega
\intsum{p}
\frac{ \braOket{q}{z}{p} \braOket{p}{z}{a} }
{\epsilon_a+\omega-\epsilon_p},
\label{MqOoa}
\end{equation}
and the integral-sum is {\it real}. 
We have found that this `reversed time-order' is negligible for monochromatic fields.
The work by Ishikawa and Ueda implies that 
phase effects will arise if large bandwiths are included in the calculation \cite{IshikawaPRL2012,IshikawaAS2013}.
However, such bound excitations can only be substantial if the first photon is close to a resonance of the system,
which is not the case for incident IR photons ($\sim 1$\,eV) on noble gas atoms ($\sim10$\,eV excitation energy). 
The reversed time-order will not be included in the following analysis.

\subsubsection{Extraction of atomic delays}
\label{sec_extractdelays}

We now turn to the extraction of the atomic delays in laser-assisted photoionization from two-photon matrix elements in the context of RABITT-type measurements. The probability of the $2N$:th sideband is modulated as a function of delay between the XUV and IR fields as:
\begin{equation}
W_{2N}=\alpha_{2N}+\beta_{2N}\cos[2\omega(\tau-\tau_\mathrm{GD}-\tau_\mathrm{A})],
\label{rabitteq}
\end{equation}
where $\alpha_{2N}$ and $\beta_{2N}$ are real numbers;  
$\tau=\varphi/\omega$ is the phase-delay of the IR field: 
$E_\mathrm{IR}(t)=2E_\mathrm{1}^{(0)}\sin[\omega (t -\tau)]$;
and $\tau_\mathrm{GD}=\tau_\mathrm{GD}(2N\omega)=(\phi_{2N+1}-\phi_{2N-1})/2\omega$ is the corresponding finite difference group delay of the XUV harmonics: 
$E_\mathrm{XUV}(t)=\sum_N E_{2N+1}^{(0)}\exp[i\phi_{2N+1}-i(2N+1)\omega t]$; and 
$\tau_\mathrm{A}$ is the atomic delay which arises from the phase difference of atomic two-photon matrix elements. 
The group delay of the attosecond pulses determine the ejection time of the photoelectron within each half period of the IR field. The IR field then acts on the photoelectron and the largest change in trajectory, which leads to the largest population of the sideband, is expected when the electron is ejected at the maximum of the IR vector potential, 
i.e. when the photoelectron is ejected in zero electric field. 
In more detail, the delay of the modulation in the sideband is also affected by the atomic delay, $\tau_\mathrm{A}$,
which arises from the interactions between electron, ion and IR field.   
For details about the RABITT method, 
we refer the reader to the pioneering work of Toma and Muller \cite{TomaJPB2002}. 
While the following discussion is focused on atomic delays from the RABITT method, the results are expected to hold also for (I)PROOF \cite{ChiniOE2010,LaurentOE2013} and attosecond streak-camera methods \cite{ItataniPRL2002,MairessePRA2005} provided that the field strengths are sufficiently weak \cite{DahlstromCP2013}. 
However, in RABITT measurements atomic delays can be computed for either angle-integrated photoelectrons or angle-resolved photoelectrons \cite{TomaJPB2002}. The advantage of angular-integrated (energy) detection is a stronger experimental signal. This is in contrast to both (I)PROOF and the streak-camera method where photoelectrons must be detected using (angular-resolved) momentum detectors.
{\it Angular-resolved} atomic delays are obtained using momentum states as final states, while {\it angular-integrated} atomic delays are found by incoherent addition of all the possible partial wave states of a certain continuum energy \cite{TomaJPB2002}. The question then arises: will these two detection schemes provide the same information about the ionization process?

An uncorrelated final state for the photoelectron with asymptotic momentum $\mathbf{k}$ can be conveniently expressed on a partial-wave basis 
\cite{DahlstromCP2013},  
%[no spin here only space]
%
\begin{equation}
\varphi_{\mathbf{k}}(\mathbf{r})=
\braket{\mathbf{r}}{\mathbf{k}}=
(8\pi)^{3/2}\sum_{L,M}i^{L}e^{-i(\sigma_{k,L}+\delta_{k,L})}
Y^*_{L,M}(\mathbf{\hat k})
Y_{L,M}(\mathbf{\hat r})R_{k,L}(r).
\label{final}
\end{equation}  
Assuming for simplicity that there are only two XUV frequencies, 
$\Omega_>$ and $\Omega_<$, 
and that they are separated by exactly two IR photons, $\Omega_>-\Omega_<=2\omega$. 
As indicated in Fig.~\ref{fig_photonpicture}~(a), 
the interference signal will involve absorption of the smaller XUV photon plus an IR photon and absorption of the larger XUV photon with emission of an IR photon. 
We set the spectral phase of the XUV fields to zero 
and give the IR field a relative phase shift, $\varphi$.
The probability  for the photoelectron to reach the final `sideband' state, $\mathbf{k}_q$, is then 
%[where I've introduced a factor 2 for spin]
\begin{equation}
W_{\mathbf{k}_q}(\varphi)=
2\sum_{m_a}\Big|
%\sum_{L_q}i^{-L_q}e^{i(\sigma_q+\delta_q)}
%Y_{L_q,m_a}(\mathbf{\hat k})\left[
%M(q,\omega,\Omega_<,a)+
%M(q,-\omega,\Omega_>,a)
S_\mathrm{emi}\mathbf{(k}_q,a)e^{-i\varphi} +
S_\mathrm{abs}\mathbf{(k}_q,a)e^{i\varphi} 
%\right]
\Big|^2,
\label{sideband_momentum}
\end{equation}
where the total contribution 
from the quantum path with emission of an IR photon is
\begin{equation}
S_\mathrm{emi}(\mathbf{k}_q,a)=
(8\pi)^{3/2}
\sum_{L_q}i^{-L_q}e^{i(\sigma_q+\delta_q)}
Y_{L_q,m_a}(\mathbf{\hat k}_q)M(q,-\omega,\Omega_>,a)
\label{Semi}
\end{equation}
with $M(q,\omega,\Omega_<,a)$ being the partial wave two-photon matrix elements from the initial state $a$ (labels $n_a,\, \ell_a, \, m_a$) to the final state $q$ (labels $k_q, \, L_q, \, M_q$), from Eq.~(\ref{MqoOa}). 
Similarly, the total contribution for the quantum path with absorption of an IR photon is
\begin{equation}
S_\mathrm{abs}(\mathbf{k}_q,a)=
(8\pi)^{3/2}
\sum_{L_q}i^{-L_q}e^{i(\sigma_q+\delta_q)}
Y_{L_q,m_a}(\mathbf{\hat k}_q)M(q,\omega,\Omega_<,a).
\label{Sabs}
\end{equation}
Note that the spectral phase-shift of the IR field is negative in the  emission processes in Eq.~(\ref{sideband_momentum}) due to complex conjugation of the field amplitude. 
%Considering two-photon processes from either $s$ or $p$ orbitals, the only allowed final angular momentum are $L_q=\{\ell_a,\, \ell_a+2\}$. 
For linear polarized fields the $m_a$ quantum number is conserved from the initial state, but an incoherent sum over all initial states of the atom,
$-\ell_a\le m_a \le\ell_a$, 
must be performed in Eq.~(\ref{sideband_momentum}),
where also the prefactor of two comes from spin degrees of freedom.
The atomic delay is found by locating the maximum of the sideband oscillation 
\begin{equation}
\tau(\mathbf{k_q},a;\omega)\equiv
\frac{\varphi_\mathrm{max}}{\omega}=\frac{1}{2\omega}\arg\left[
\sum_{m_a} S_\mathrm{emi}(\mathbf{k_q},a)S_\mathrm{abs}^*(\mathbf{k_q},a)
\right],
\label{taumomentum}
\end{equation}
where we emphasize the dependence on the final momentum of the photoelectron, the  orbital being photoionized and the energy of the IR photon used to probe the process.
In the special case of detection along the field polarization axis,
$\mathbf{\hat k_q}=\mathbf{\hat z}$, 
only the $m_a=0$ term in Eq.~(\ref{taumomentum}) will contribute due to properties of spherical harmonics.  

The angle-integrated probability can be calculated by integration over all ejection angles in 
Eq.~(\ref{sideband_momentum}), 
but it is more convenient to sum incoherently the probabilities of all allowed final partial-wave states with the correct final energy, $E_q$, 
%[where I've introduced a factor 2 for spin]
\begin{equation}
W_{E_q}(\varphi)=2\sum_{L_q,m_a}\Big|
\underbrace{
M(q,\omega,\Omega_<,a)e^{i\varphi}
}_{\mathrm{absorption}}
+
\underbrace{
M(q,-\omega,\Omega_>,a)e^{-i\varphi}
}_{\mathrm{emission}}
\Big|^2,
\label{sideband_energy}
\end{equation}
where the first term in the bracket is absorption of an IR photon and the second term is emission of an IR photon leading to the same partial wave state, $q$.
The atomic delay is again found by locating the maximum of the sideband oscillation 
\begin{equation}
\tau(E_q;n_a,\ell_a;\omega)\equiv
\frac{\varphi_\mathrm{max}}{\omega}=
\frac{1}{2\omega}\arg\left[
\sum_{L_q,m_a}
M(q,-\omega,\Omega_>,a)
M^*(q,\omega,\Omega_<,a)
\right].
\label{tauenergy}
\end{equation}
This shows that if there is only one (dominant) final partial wave state, then the angle resolved delay, Eq.~(\ref{taumomentum}), and the energy integrated delay, Eq.~(\ref{tauenergy}), will be equivalent. 
However, in the general case the atomic delays may differ.

\subsection{Numerical implementation}
\label{sec_numint}
There are several issues to discuss in connection with the choice of basis functions. 
The first issue is the choice of
potential;  we have used a Hartee-Fock (HF) potential with a correction that provides an asymptotically correct long-range interaction for the photoelectrons in Eq.~(\ref{finalasymp}). A second point is the numerical representation of the basis function; which is here done with so-called B-splines. Finally,
the construction of an outgoing wave packet requires integration over the pole in Eq.~(\ref{twophotonpert}). A numerical stable and efficient way to do this is provided by the method of complex scaling, which we here use in the form of exterior complex scaling 
\cite{NicolaidesPLA1978,SimonPLA1979}.

\subsubsection{Atomic potential and basis}
Starting with the first issue, we carry out  the perturbation expansion with partial wave states
$P_{n\ell}(r) Y_{\ell m}(\theta,\phi)$, 
where  $P_{n\ell} \equiv rR_{n\ell}$ 
are eigenstates with eigenvalues $\epsilon_{n,\ell}$ 
(for simplicity now representing  both bound and continuum states) 
to the effective one-particle Hamiltonian:
\begin{equation}
h_{\ell} = -
%\frac{\hbar^2}{2m} 
\frac{1}{2} 
\frac{\partial^2}{\partial r^2} 
+
%\frac{\hbar^2}{2m} 
\frac{1}{2}
\frac{\ell (\ell + 1)}{r^2} 
- 
%\frac{e^2}{4\pi \varepsilon_0} 
\frac{Z}{r} + u_\mathrm{HF} + u_\mathrm{proj}.
\label{h_hf}
\end{equation}
The Hartree-Fock potential, $u_\mathrm{HF}$, 
\begin{equation}
\braOket{c}{u_\mathrm{HF}}{a} = \sum_b^{core} 
\braOket{c b}{
%\frac{e^2}{4\pi \varepsilon_0} 
\frac{1}{r_{12}}}{ab}
- 
\braOket{bc}{
%\frac{e^2}{4\pi \varepsilon_0} 
\frac{1}{r_{12}}}{ab}
\label{u_hf}
\end{equation}
accounts for the bulk of the interaction between the closed shell ground state electrons. The first term on right-hand side of Eq.~(\ref{u_hf}) is the local Hartree potential, while the second term is the non-local exchange potential. 
In addition we include a potential from a core hole, $u_\mathrm{proj}$, that
 ensures a good starting point 
for excited electrons without affecting the core electrons.
 It is constructed with projection operators
\begin{equation}
 \sum_{p}^{exc} \ket{p}\bra{p}  = 
1 - \sum_{b}^{core}  
 \ket{b}\bra{b}
\end{equation}
to form a so-called projected potential 
\begin{equation}
u_\mathrm{proj}  = -\sum_{p,q}^{exc}  
\ket{p}
%\underbrace{
\braOket{pv}
{ 
%\frac{e^2}{4\pi \varepsilon_0} 
\frac{1}{r_{>}}
}
{q v}
%}_{\mathrm{Monopole \ only}}
\bra{ q},
\label{u_proj}
\end{equation}
where $v$ denotes the removed core-orbital and $1/r_>$ is the monopole term in the multipole expansion of $1/{r_{12}}$ \cite{mbpt}. The projected potential attracts the excited electrons so that the  basis states will include not only continuum energies with positive energy, $\epsilon_{k,\ell}>0$; but also bound excited orbitals with negative energy, $\epsilon_{n,\ell}<0$. The number of bound orbitals depends on the design of the B-spline knot-sequence, discussed in Sec.~\ref{sec_bspline}.
Including higher-order terms in the multipole expansion of the projected potential in Eq.~(\ref{u_proj}) leads to short-range corrections of the atomic potential. These interactions, together with true correlation effects, will be accounted for using the diagrammatic expansion of the photoelectron wave packet in Sec.~\ref{sec_diagcorr}. 

\subsubsection{B-spline representation}
\label{sec_bspline}
Turning to the numerical representation of the basis function, we  expand the  radial functions $P_{n\ell}(r)$ in  B-splines \cite{deboor},
\begin{equation}
P_{n\ell} (r) =    \sum_i c_i B_i^k (r),
\label{Bexp}
\end{equation}
which are piecewise polynomials of order $k-1$, defined on a so-called knot sequence. The B-splines form a finite basis that is complete on the space determined by the polynomial order and the knot sequence.  
The expansion coefficients, $c_i$, in Eq.~(\ref{Bexp}) are obtained from a
diagonalization of the one-particle Hamiltonian, Eq.~(\ref{h_hf}),  in the B-spline basis.
The hereby obtained functions, $P_{n\ell}$, constitute an orthonormal finite basis  where a certain number of states coincide with the physical eigenstates to Eq.~(\ref{h_hf}) and the rest are forming a pseudo-continuum that can be used to express the perturbed wave function.
Since both $u_\mathrm{HF}$ and $u_\mathrm{proj}$ depend themselves on the the eigenstates of Eq.~(\ref{h_hf}) the $c_i$ coefficients in Eq.~(\ref{Bexp}) are determined through successive diagonalizations of the Hamiltonian matrix  in an iterative scheme.

Finally, exterior complex scaling is introduced through 
a transformation of the radial variable as \cite{NicolaidesPLA1978,SimonPLA1979}
\begin{align}
r\rightarrow 
\left\{
\begin{array}{lr}
r, 	& 			 	0<r<R_C	\\
R_C + (r-R_C)e^{i\varphi}, & 	R_C<r.
\end{array}
\right. 
\label{ECS}
\end{align}
The transformation is in  practice made  by 
 letting the
 knot sequence, defining the B-splines, follow 
 $r$ in Eq.~(\ref{ECS}),
with one knot exactly at $R_C$.
Our implementation follows mainly the description in~\cite{mccurdy:extcs:2004}.
Exterior complex scaling has several desirable properties: The eigenenergies to the 
complex-scaled Hamiltonian, i.e.  
Eq.~(\ref{h_hf}) with $r$ transformed as in Eq.~(\ref{ECS}), are complex and 
the integration over the  poles in Eq.(\ref{twophotonpert}) can then be replaced with 
a summation of pseudo-continuum states, which effectively performs an integration along a path in the complex plane. Still the perturbed function in the unscaled region is unaffected 
by the scaling and with a large enough  $R_C$, the perturbed wave function can thus  be analysed 
in a region well outside the bound electrons. Our knot sequence is  linear in most of the unscaled region, with a step of typically 0.1-0.3 a.u. 
%(Check! I use 0.25) 
between the  knots. This
is to ensure  
a good description of the continuum. Close to the nucleus there are a few extra knot points 
for the representation of the inner electrons, and outside $R_C$, where the wave function is damped, a less dense grid can be used.

\subsection{Diagrams for correlated interactions}
\label{sec_diagcorr}
While one-color two-photon ionization of atomic systems has been studied for many years, 
c.f.~Refs.~
\cite{HuillierPRA1987,KobayashiOL1998,MoshammerOE2011},  
two-color ionization has only attracted interested more recently due to the development of attosecond metrology using coherent XUV and IR fields. 
Prior to our work \cite{DahlstromPRA2012}, however, no studies of correlated atomic systems, beyond helium, were carried out, with the exception of a time-dependent R-matrix calculation in neon by Moore et al. \cite{MoorePRA2011}. In the following, we give details on how correlation is included in our two-color calculation. 

\subsubsection{Screening of the XUV photon by correlation effects}

\begin{figure}[ht]
	\centering
	\includegraphics[width=\textwidth]{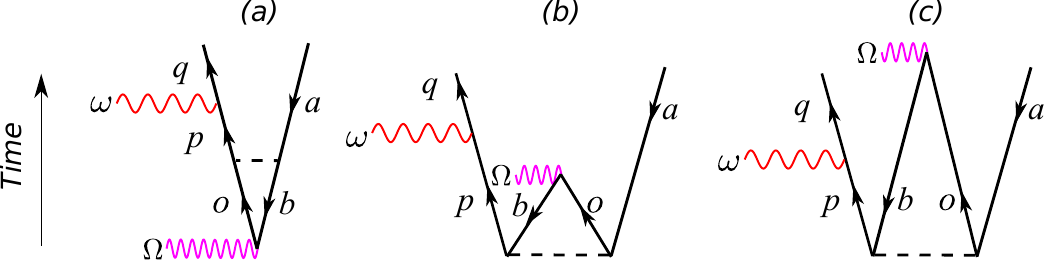}
	\caption{\label{fig_basicdiagrams} 
(Color online)
Two-photon absorption processes: (a) Direct Coulomb interaction in-between absorption of photons; (b) Ground-state correlation where IR photon is absorbed last; (c) Ground-state correlation where XUV photon is absorbed last. For details about notation see main text and Ref.~\cite{mbpt}. 
} 	
\end{figure}

In order to include correlated interaction into the photoionization process, we consider the perturbation
\begin{equation}
\delta V=
% \frac{e^2}{4\pi \varepsilon_0} 
\frac{1}{r_{12}} - u_\mathrm{HF}-u_\mathrm{proj},
\label{perturbation}
\end{equation}
indicated as dashed lines in Fig.~\ref{fig_basicdiagrams}. 
In Fig.~\ref{fig_basicdiagrams}~(a) we include a single direct interaction after absorption of the first photon but before absorption of the second photon. The photons are drawn as waves while the electron and holes are drawn as up and down arrows, respectively.  This interaction leads to a first-order correction of the intermediate electron wave packet due to correlation with the ion. Note that the state of the hole may be changed in this process and that all possible intermediate angular momenta and magnetic quantum numbers must be taken into account. 
The inclusion of $u_\mathrm{HF}$ in Eq.~(\ref{perturbation}), rather than the full Coulomb matrix element, cancels classes of HF diagrams \cite{mbpt}. The use of $u_\mathrm{proj}$ ensures that the excited states are Coulomb-type, but it also requires that certain diagram contributions are skipped. As an example, the diagram shown in Fig.~\ref{fig_basicdiagrams}~(a) is skipped for $a=b=v$ with the monopole interaction, as it is already `included' in the basis. 
The perturbation matrix element for the direct Coulomb interaction in so-called  `forward' propagation [Fig.~\ref{fig_basicdiagrams}(a)] is 
\begin{equation}
M^{(1)}_{\mathrm{fwd}}=
-\sum_{p,o}^{\mathrm{exc.}}\sum_{b}^{\mathrm{core}}
\frac{ \braOket{q}{z}{p}\braOket{pb}{\delta V}{oa}\braOket{o}{z}{b} }
{(\epsilon_a+\Omega-\epsilon_p)(\epsilon_b+\Omega-\epsilon_o)},
\label{basicdiagrambeq}
\end{equation}
corresponding to absorption of the XUV photon at any time before the Coulomb interaction.

The lowest-order absorption process can also occur as a result to ground-state correlation in the initial atomic state, where {one} {\it virtual} electron--hole pair is first created and then annihilated by absorption of an XUV photon. 
As shown in Fig.~\ref{fig_basicdiagrams}~(b) and (c), this annihilation may occur before or after absorption of the IR photon, respectively. 
Note that the interaction structure in 
Fig.~\ref{fig_basicdiagrams}~(b) and (c)
are identical and that the only difference between the two diagrams is 
found in the propagation factors (denominators). 
By addition of both factors we obtain:
\begin{equation}
\frac{1}{
(\epsilon_a+\Omega-\epsilon_p)
(\epsilon_a+\epsilon_b-\epsilon_p-\epsilon_o)
}
+
\frac{1}{
(\epsilon_a+\epsilon_b+\omega-\epsilon_q-\epsilon_o)
(\epsilon_a+\epsilon_b-\epsilon_p-\epsilon_o)
}
\nonumber
\end{equation}
\begin{equation}
=\frac{1}{
(\epsilon_a+\Omega-\epsilon_q)
(\epsilon_b-\Omega-\epsilon_o)},
\label{addgroundstatecorr}
\end{equation}
where we used energy conservation, 
$\epsilon_q-\epsilon_a=\Omega+\omega$.
The combined two time-orders becomes 
the so-called {\it reverse} propagation term
\begin{equation}
M^{(1)}_{\mathrm{bwd}}=
-\sum_{p,o}^{\mathrm{exc.}}\sum_{b}^{\mathrm{core}}
\frac{ \braOket{q}{z}{p}\braOket{po}{\delta V}{ba}\braOket{o}{z}{b} }
{(\epsilon_a+\Omega-\epsilon_p)(\epsilon_b-\Omega-\epsilon_o)},
\label{basicdiagramb}
\end{equation}
which corresponds to absorption of the XUV photon 
at any time {\it after} the Coulomb interaction. 
Note that the sign on the photon energy in the second denominator is negative. 
This minus sign removes the divergence from 
the excited spectra because: 
$\epsilon_o>\epsilon_b>\epsilon_b-\Omega$, and the factor will
monotonically decrease with increasing excited state energy, 
$\epsilon_o$.
In this some sense,  
the electron-hole pair 
first travels backward in time 
(the frequency of the photon is felt with a minus sign)
and then, after the Coulomb interaction, 
it propagates forward in time (positive sign of photon energy because the photon has been absorbed).

\begin{figure}[ht]
	\centering
	\includegraphics[width=\textwidth]{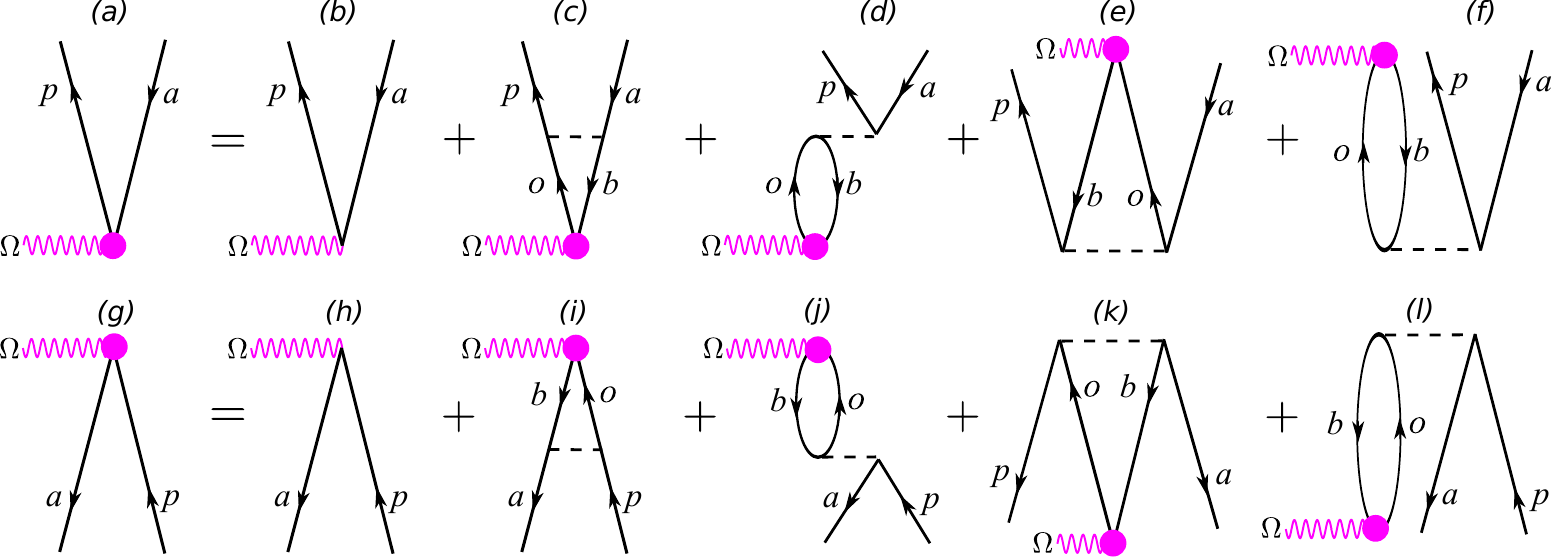}
	\caption{\label{fig_fforwardbbackward} 
(Color online)
Random-phase approximation with exchange (RPAE) for many-body screening effect of the XUV photon. (a) and (g) are forward and reverse propagation, respectively, where the sphere indicates correlated interaction to infinite order. For details see Ref.~\cite{Amusia1990}.  
} 	
\end{figure}

In Fig.~\ref{fig_fforwardbbackward} we show the generalized  integral equation 
for the `infinite order' perturbed wavefunction including also Coulomb exchange interactions 
for forward Fig.~\ref{fig_fforwardbbackward}~(a)--(f) 
and reverse propagation (g)--(l). 
Interestingly, these equations are identical to 
the one-photon RPAE method \cite{Amusia1990}.
The forward correlated perturbed wavefunction, Fig.~\ref{fig_fforwardbbackward}~(a), should then be substituted into the
uncorrelated two-photon matrix element, Fig.~\ref{fig_photonpicture}~(b). 
Our method for evaluation of the IR transition 
is explained in Sec.~\ref{sec_IRinducedtrans}. 

\subsubsection{Further screening effects}
A natural extension of this work is to include interactions with $\delta V$ from Eq.~(\ref{perturbation})
also {\it after} absorption of the IR photon,
i.e. to go beyond the monopole interaction approximation 
with $v$ for the basis states given in Eqs.~(\ref{h_hf})  and (\ref{u_proj}), 
and consider a correlated final state. 
In RPAE-language this corresponds to a linear screening of two photons of different color: $\Omega+\omega$.
Such linear screening effects are small 
because they rely mainly on quadropole interactions \cite{HuillierPRA1987}. 
Further, the absolute scattering phase of the final state cancels in uncorrelated attosecond delay experiments \cite{DahlstromCP2013}, so the monopole term in $u_\mathrm{proj}$ should be quite sufficient for our method. 
Non-linear screening effects in two-photon absorption, 
also introduced in Ref.~\cite{HuillierPRA1987}, 
are probably negligible in laser-assisted photoionization because the IR photon 
does not have enough energy to excite an additional electron-hole pair 
(see Sec.~\ref{sec_irinit}).

\subsubsection{Autoionizing resonances}
\label{sec_res}
\begin{figure}[ht]
	\centering
	\includegraphics[width=\textwidth]{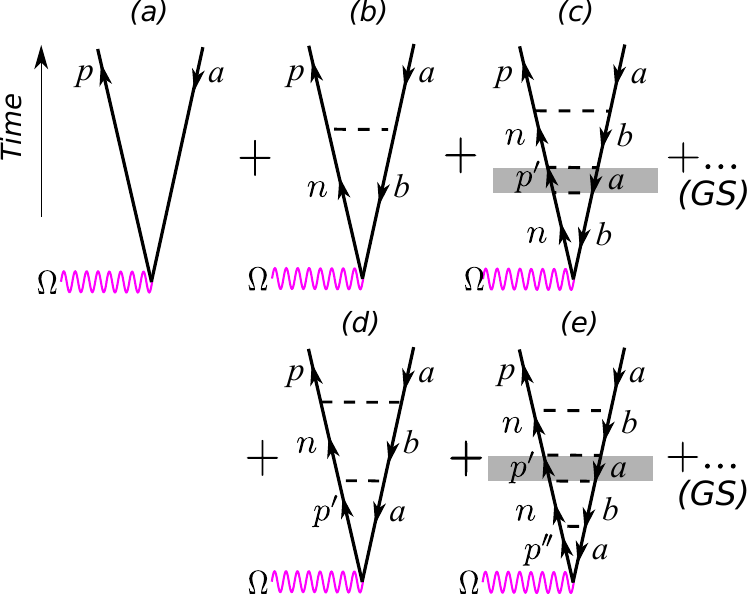}
	\caption{\label{fig_fano_diagrams} 
(Color online)
Perturbation diagrams for 
photoionization including 
interaction with a single resonance. 
The processes are:
(a) direct ionization process; 
(b) dipole excitation to the resonance and then decay to the continuum;
(c) correction to (b) with a virtual transition to the continuum (grey box);
(d) direct ionization with a virtual transition to the resonance;
(e) correction to (d) with a virtual transition to the continuum (grey box). 
The perturbation series (b)+(c)+... and (d)+(e)+... 
can be extended as a geometric series (GS) 
given in Eq.~(\ref{fanogeom}). 
} 	
\end{figure}

Atoms in excited states with energy above the first ionization threshold can decay through correlation effects, 
as first explained by Fano in Ref.~\cite{fanoPR1961}. The general theory of autoionizing resonances  was developed using configuration interaction  
\cite{fanoPR1961,staracePRA1977}, 
but the evaluation can also be 
carried out using infinite-order perturbation theory, 
e.g. RPAE \cite{Amusia1990}.
In Fig.~\ref{fig_fano_diagrams} we show lowest-order perturbation diagrams
that account for coupling of an inner hole, $b$, and an outer hole, $a$. 
The electron-hole pair will pass through   
the resonance with bound electron energy: $\epsilon_n=\epsilon_b+\Omega<0$, 
to the final photoelectron state with energy: $\epsilon_p=\epsilon_a+\Omega>0$. 
The diagrams in Fig.~\ref{fig_fano_diagrams} represent a subset of RPAE diagrams that can be continued as geometric series (GS) and, therefore, evaluated to infinite-order analytically.
In this subsection we consider one resonance and one continuum for simplicity. The result of this procedure leads to an effective dipole matrix element
\begin{equation}
Z=
z+
\frac{c^*}{\epsilon_b+\Omega-\epsilon_n-\Delta}
\Big(
-
%\mp
z_R
+\lim_{\varepsilon\rightarrow0^+}
\intsum{p'}\frac{c'z'}{\epsilon_a+\Omega-\epsilon_{p'}+i\varepsilon}
\Big), 
\label{fanogeom}
\end{equation}
where the uncoupled dipole matrix element to the continuum and  to the resonance is $z=\braOket{p}{z}{a}$ and $z_R=\braOket{n}{z}{b}$, respectively; 
and the direct Coulomb matrix element is $c=\bracorrket{pb}{na}$. 
Primes indicate that $p$ is replaced by $p'$ in the matrix elements. 
The minus sign in front of $z_R$ follows from the Goldstone rules, c.f. Ref.~\cite{mbpt}. 
The shift and broadening of the resonance in Eq.~(\ref{fanogeom}) is:  
\begin{eqnarray}
\Delta&=&\Delta_R+i\Delta_I \nonumber \\
&=&
\lim_{\varepsilon\rightarrow 0^+}
\intsum{p'}\frac{\left|c'\right|^2}
{\epsilon_a+\Omega-\epsilon_{p'}+i\varepsilon}
\nonumber \\
&=&
\mathrm{p.v.}
\intsum{p'}\frac{\left|c'\right|^2}
{\epsilon_a+\Omega-\epsilon_{p'}}
-i\pi
\left|c\right|^2
,
\label{complexshift}
\end{eqnarray}
where $|c'|^2=|\braOket{p'b}{r^{-1}_{12}}{na}|^2$ is the coupling strength; 
and  $p$ (without prime) is the on-shell state, $\epsilon_p=\epsilon_a+\Omega$. 
The effective dipole matrix element in 
Eq.~(\ref{fanogeom}) can be written on the famous Fano form \cite{fanoPR1961}:
\begin{equation}
\frac{Z}{z}=
\frac{(\epsilon+q)}{\epsilon+i}=
\frac{(\epsilon+q)(\epsilon-i)}{\epsilon^2+1},
\label{fanoformula}
\end{equation}
using the energy parameter:  
$\epsilon=(\epsilon_b+\Omega-\epsilon_n)/\pi|c|^2$, 
and the $q$-parameter:
\begin{equation}
q=
\frac{1}{\pi c z}
\Big(
%\mp 
-z_R+
\mathrm{p.v.}\intsum{p'}\frac{c'z'}{\epsilon_a+\Omega-\epsilon_p'}
\Big).
\end{equation}
Using Eq.~(\ref{fanoformula}), we identify that 
the phase of the one-photon dipole matrix element acquires a smooth positive $\pi$-shift over the resonance (close to $\epsilon=0$) 
but also an abrupt $\pi$-shift ($\epsilon=-q$): 
\begin{equation}
\Phi_R(\Omega)=
\arg\left(\frac{Z}{z}\right)=
\mathrm{atan}\left(\frac{-1}{\epsilon}\right)+\pi\Theta(-\epsilon-q)
%=
%\mathrm{atan}\left(\frac{-\pi|c|^2}{\epsilon_b+\Omega-\epsilon_n}\right)
,
\label{fano_phase}
\end{equation} 
where $\Theta(-\epsilon-q)$ is the Heaviside stepfunction (which is zero for $\epsilon>-q$).
Far from the resonance, the phase is unchanged by the resonance 
(modulo $2\pi$).  
Closer to the resonance, the phase is a function of the incident XUV photon energy and the delay of the photoelectron wave packet becomes a positive Lorentzian curve plus a sharp spike: 
\begin{equation}
\tau_R(\Omega)=
\frac{d\Phi_R}{d\Omega}=
\frac{\pi|c|^2}
{(\epsilon_b+\Omega-\epsilon_n)^2+(\pi|c|^2)^2}+\frac{d}{d\Omega}\pi\Theta(-\epsilon-q).
\label{resdelay}
\end{equation}
We note that the sharp spike occurs in a region where the photoionization cross-section exactly vanishes. 
Provided that the photoelectron wave packet does not overlap with this minima,  
Eq.~(\ref{resdelay}) implies that the energy components of the electron wave packet will be {\it delayed} by a finite amount due to the resonance. The largest delay occurs at the original position of the uncorrelated resonance:
$\epsilon=\epsilon_b+\Omega-\epsilon_n=0$ and it scales with the inverse coupling strength. 
The electron will remain close to the atom in a quasi-stable state for a longer time if the coupling out to the continuum is weak. 
We stress that attosecond metrology, 
based on interferometry as shown in 
Fig.~\ref{fig_photonpicture}~(a),
is likely to ``fail'' when the IR photon is larger 
than the typical spectral extent of a resonance. 
This is due to the break down of the finite-difference approximation to the spectral derivative, 
i.e. to the Wigner delay of the photoelectron.
Under the na\"ive assumption that there is only {\it one} autoionizing resonance, and that all other phase effects can be negelected, the measured delay using RABITT is expected to become: 
\begin{equation}
\tau_R^\mathrm{(RABITT)}(\Omega)=
%\frac{\Phi_R(\Omega+\omega)-\Phi_R(\Omega-\omega)}{2\omega}=
\frac{\Phi_R(\Omega_>)-\Phi_R(\Omega_<)}
{2\omega},
\end{equation}
where 
the first  term comes from the emission path (right) and 
the second term comes from the absorption path (center) 
in Fig.~\ref{fig_photonpicture}~(a). 
Assuming that the resonance is narrow 
compared to the IR photon energy, 
$\omega\gg\pi|c|^2$, 
only {\it one} of the two quantum paths will pass close to the resonance 
and experience a phase shift.
The observed delay simplifies to: 
$\tau_R\approx\Phi_R(\Omega_>)/2\omega$ or 
$\tau_R\approx-\Phi_R(\Omega_<)/2\omega$, 
corresponding to the case when the emission or absorption path is closest to the resonance. For a simple illustration, see Fig.~\ref{fig_Ar_fano}~(b). 
Rather than recording the delay of the resonance, it is the {\it phase} of the resonance that will be recorded using the RABITT scheme. We note that the phase will be recorded with a scaling factor of $1/2\omega$ with either a positive or negative sign depending on which quantum path is closest to the resonance.   

%we may replace Eq.~(\ref{fano_phase}) by a Heavyside step-function, 
%$\Phi_R(\Omega)\approx\pi\Theta(\Omega-\epsilon_R)$, 
%where $\epsilon_R=\epsilon_n-\epsilon_b$.
%As the XUV frequency is tuned the delay will first increase by a quarter of the IR period, 
%$\pi/(2\omega)=T_\mathrm{IR}/4$, 
%at 
%$\Omega_>=\epsilon_R$,
%then it will stay constant until 
%$\Omega_<=\Omega_>-2\omega=\epsilon_R$, 
%where it will return to zero.
%Indeed, the observed `box-shaped delay' is a poor approximation to the Lorentzian delay peak.
%The situation gets worse if there are multiple narrow resonances leading to multiple jumps of $T_\mathrm{IR}/4$ before returning to the nominal value. Problems arise because the RABITT measurement made in modulus $T_\mathrm{IR}/2$, which implies that two narrow resonances may cause the illusion of no delay! In a realistic situation, such as an autoionizing Rydberg series, there will be multiple  resonances in the energy region between the two XUV harmonics, which leads to a complications for interferometric delay measurements. 
 
As already mentioned, the RPAE includes infinitely many more diagrams, but, still, only `single'--type autoionizing resonances are included. It has been shown by Amusia and Kheifets that complicated double excitations beyond the RPAE are required for good agreement with experimental Fano parameters of the $3s^{-1}np$-autoionization resonances in argon \cite{Amusia1981,Amusia1982}. 
More recently, the quantitative role of double excited states in atomic delay measurements were explored by Carette et al. using multiconfigurational Hartree-Fock \cite{CarettePRA2013}. Our diagrammatic method can be extended to include also double-excitation resonances, but this is beyond the scope of the present paper where we focus of RPAE correlation effects.

\subsection{IR-induced continuum transition}
\label{sec_IRinducedtrans}

The asymptotic Coulomb properties of the wavefunctions have been 
shown to be important for the phase of the two-photon matrix element 
\cite{AymarJPB1980,TomaJPB2002,DahlstromCP2013}. 
The radial integral on the right-hand side of 
Eq.~(\ref{twophotonpert}) is therefore divided into two parts: 
An inner region $0\le r<R$, 
where the perturbed wavefunction and final state 
can be determined numerically on the B-spline basis; 
and an outer region $R\le r<\infty$ 
where the functions 
behave approximately as shown in
Eqs.~(\ref{finalasymp}) and (\ref{pertasymp}). 
The phase-shifts of the perturbed wavefunction and of the final state, 
$\delta_p$ and $\delta_q$, 
are determined by a matching procedure to 
the numerical wavefunctions at $r=R<R_\mathrm{ecs}$. 
The outer part of the dipole integral extends to infinity 
and a straightforward numerical radial integration is, therefore, 
not possible.  

\begin{figure}
	\centering
	\includegraphics[width=\textwidth]{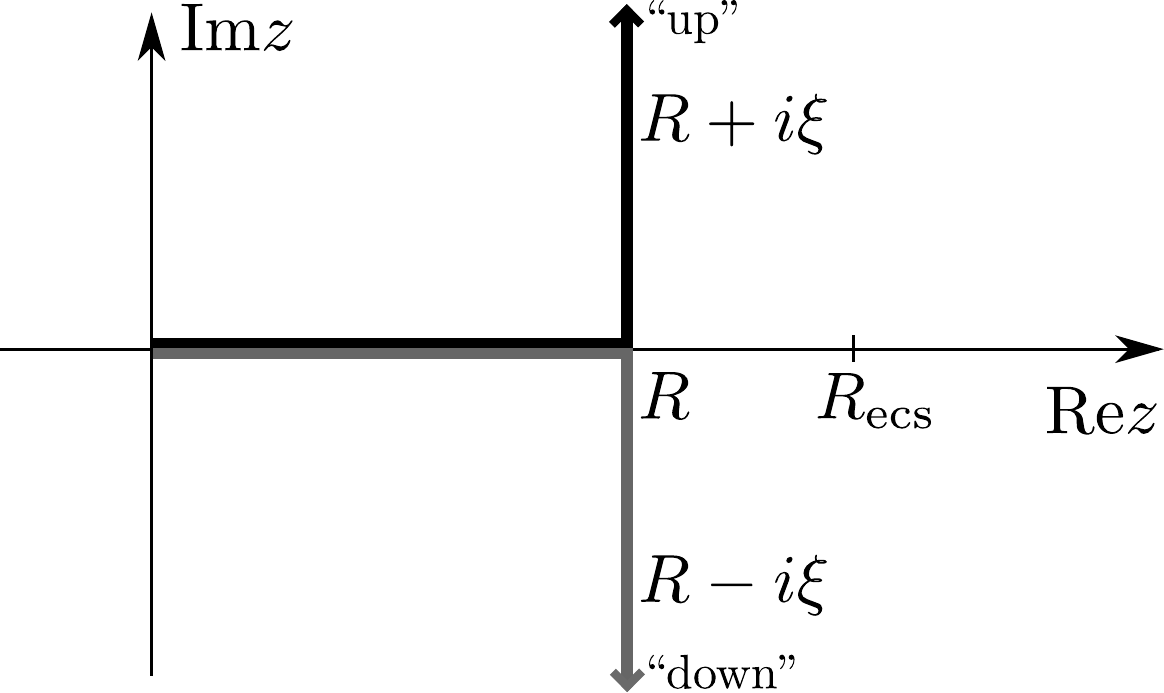}
	\caption{\label{fig_complexplane} 
(Color online) Radial integration path in the complex plane for the dipole interaction with the IR photon between two wavefunctions of continuum character.
} 	
\end{figure}

% MATCHING PROCEDURE :
In our numerical implementation we use the \texttt{COULCC} program by Thompson and Barnett for evaluation of Coulomb functions with complex arguments \cite{coulcc1985}.
The regular and irregular Coulomb functions,  
$F_l(\eta,kr)\equiv F_{k,l}(r)$ and 
$G_l(\eta,kr)\equiv G_{k,l}(r)$, 
are of sin and cos--type, respectively.
Outgoing $(+)$ and ingoing $(-)$ Coulomb functions 
can be constructed as
\begin{equation}
F^{(\pm)}_{k,l}(r)=G_{k,l}(r)\pm iF_{k,l}(r)\approx 
\exp[\pm i(kr+k^{-1}\ln 2kr -\pi l/2 +\sigma_l)],
\label{coulomboutin}
\end{equation}
for $r\rightarrow\infty$.
The final state $q$ (with $k_q$ and $\ell_q$) is expressed in the asymptotic region as
a real standing wave, i.e. as a sum of phase-shifted outgoing and ingoing waves
\begin{equation}
\lim_{r\rightarrow\infty}
R_q(r)=
\frac{N_q}{r}
\times\frac{1}{2i}
\left[
\exp(i\delta_q)F^{(+)}_q(r)-\exp(-i\delta_q)F^{(-)}_q(r)
\right].
\label{pertcoul}
\end{equation}
In general, the energy of the final state will not match any of the basis states and we need to solve the uncorrelated Hamiltonian, Eq.~(\ref{h_hf}), for each $\ell_q$ at the final energy, $\epsilon_q$. 
In order to determine the phase-shift, $\delta_q$, from the numerical solution we apply an asymptotic expansion of the Coulomb function (see Sec.~14.5 in Ref.~\cite{abramowitz:437}). 
The reduced perturbed wavefunction,
Eq.~(\ref{pertasymp}), is an outgoing wave
\begin{equation}
\lim_{r\rightarrow\infty}
\rho_{p}(r)=
-\pi\frac{N_p}{r}
\braOketRed{p}{r}{a}\times \exp(i\delta_p)F^{(+)}_p(r),
\label{fincoul}
\end{equation}
where the dipole element should be replaced by the complex 
dipole, Fig.~\ref{fig_fforwardbbackward}~(a), in the correlated case. 
The non-Coulomb phase-shift $\delta_p$ of the outgoing perturbed wavefunction 
plus correlation corrections are extracted using Eq.~(\ref{pertasymp}) and (\ref{coulomboutin}): 
\begin{equation}
\delta_p+\arg[\braOketRed{p}{r}{a}]=\pi+\arg\left[\rho_p(R)/F^{(+)}_{p}(R)\right]
\label{extractdp}
\end{equation} 
%where we use the short-hand notation for momentum and angular momentum, $k_p$ and $\ell_p$, and 
where $R<R_\mathrm{ecs}$ is a large radial distance from the atom. 
%
%[Why this convernsion in code? ana=-ii*(g+ii*f) -- is this the 1/i factor? But then the minus is missing?]
%
The outer part of the radial IR dipole integral is deformed in the complex plane as
\begin{eqnarray}
I_\mathrm{outer}&=\int_R^{\infty} dr \ r^3 \ R_q(r) \ \rho_p(r) 
\nonumber \\
&\approx
\frac{\pi}{2i}N_qN_p\braOketRed{p}{r}{a}\times
\left[
J_\mathrm{outer}^{(-)}e^{-i\delta_q} \ - \ 
J_\mathrm{outer}^{(+)}e^{+i\delta_q}
\right]e^{i\delta_p}
\label{rintouter}
\end{eqnarray}
where two Coulomb integral terms are
\begin{equation}
J_\mathrm{outer}^{(\pm)}=
\int_{C} dz \ z F^{(\pm)}_q(z)  F^{(+)}_p(z).
\label{Jintegral}
\end{equation}
Consider the case of absorption of an IR photon, $k_q>k_p$.
The $J_\mathrm{outer}^{(+)}$ integral will then be exponentially
converged when evaluated along the contour:   
$z=R+i\xi$ for $\xi\rightarrow\infty$, 
labelled ``up'' in Fig.~\ref{fig_complexplane}.
The exponential convergence follows from the asymptotic properties 
of the outgoing Coulomb functions in Eq.~(\ref{coulomboutin}). 
The $J_\mathrm{outer}^{(-)}$ integral should instead be integrated
 ``down'' in the complex plane for convergence: 
$z=R-i\xi$ with $\xi\rightarrow\infty$. 
In practice convergence is found for finite excursions 
into the complex plane and the there is no need to integrate 
all the way to infinity. 
For the process of IR-photon emission we have $k_q<k_p$
which implies that both integrals, $J_\mathrm{outer}^{(\pm)}$, 
should be integrated ``up'' in the complex plane for convergence.
We have found good stability of this method by performing the integral 
from different values of breakpoints, $R<R_\mathrm{ecs}$. 
%

%\input{num_res.tex}

% NUMERICAL RESULTS
\section{Numerical results}
\label{sec_numres}

\subsection{General features of the atomic delay from the $M$-shell in argon}

% Ar3p w/ Muller + Schafer + RPAE (HF/EXP)
% Ar3s(C3s,Cnot1s) w/ RPAE (HF/EXP)

\begin{figure}[h]
	\centering
	\includegraphics[width=0.51\textwidth]{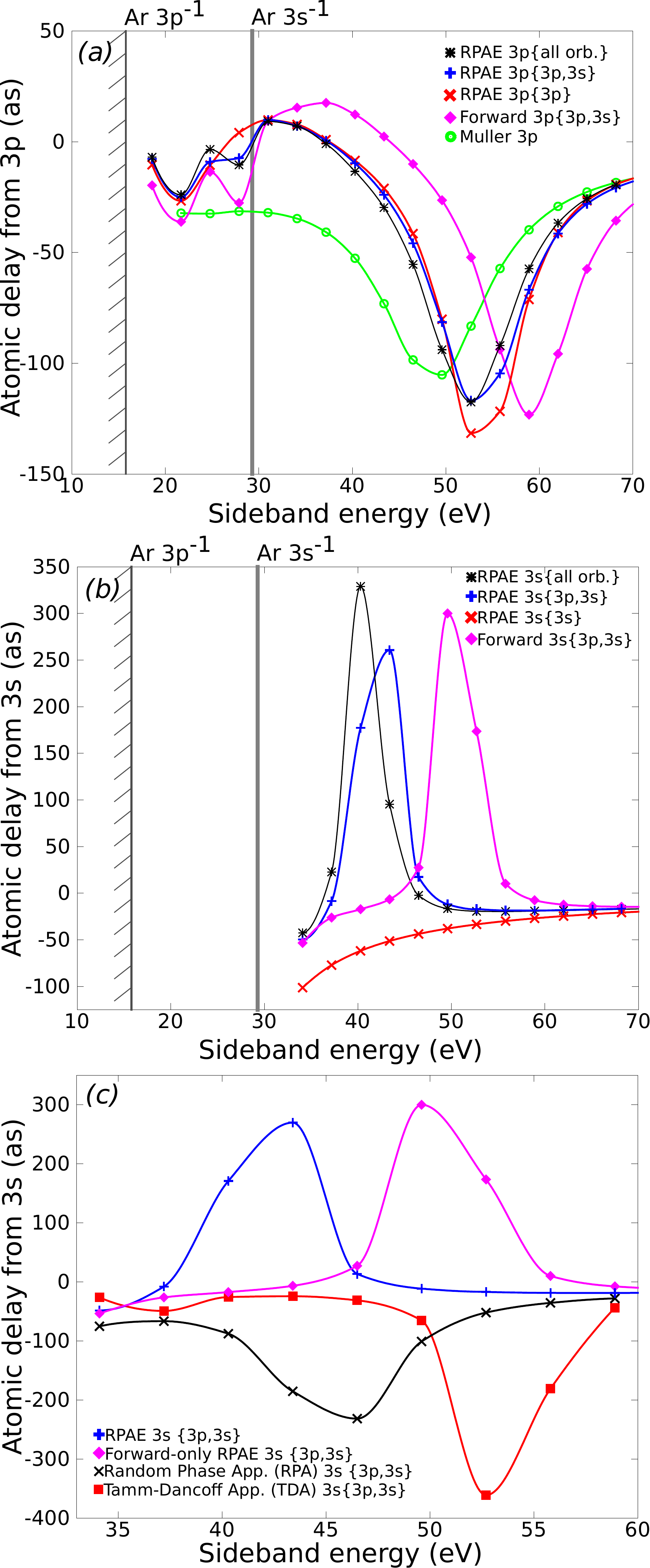}
	\caption{\label{fig_delay} 
(Color online)
Atomic delay in argon observed along the polarization axis of the fields. 
(a) and (b) correspond to ionization from 
Ar$3p^{-1}$ and Ar$3s^{-1}$, respectively. 
The correlated orbitals in the RPAE model are listed in the curly brackets.
(c) The sign of the $3s$-atomic delay peak differs between the different correlation models. 
The IR photon is $1.55$\,eV.
} 	
\end{figure}

In Fig.~\ref{fig_delay}~(a) and (b) we present the atomic delays for photoelectrons from the $3p$ and $3s$ orbitals in argon, respectively, 
with detection restricted to the polarization axis of the fields 
(see Eq.~(\ref{taumomentum}) with $\mathbf{\hat k_q}=\mathbf{\hat z}$). 
The different curves correspond to the same correlation models already introduced for the partial cross-sections in Fig.~\ref{fig_cross}. 
The data points are computed for the RABITT scheme \cite{MullerAPB2002,TomaJPB2002}, where atomic delay measurements are restricted 
to sideband energies: 
$2N\omega$ with $\omega=1.55$\,eV for integer values of $N$. 
The data points are connected by curves to guide the eye.
Before our brief report \cite{DahlstromPRA2012}, 
the only available theoretical atomic delay calculations for argon where carried out by Toma and Muller \cite{TomaJPB2002} and Mauritsson et al. \cite{MauritssonPRA2005} 
%using pseudo-potential calculations 
within the single-active electron approximation.  
The atomic delays obtained using the Muller potential, Eq.~(\ref{mullerpot}),  
were used to interpret the first experimental measurement of an attosecond pulse train \cite{PaulScience2001}.
We have verified that the atomic delays obtained using our program with the Muller potential [green $\opencircle$ in Fig.~\ref{fig_delay}~(a)] 
are in excellent agreement %(error of $2\,$as) 
with the numerical results of Ref.~\cite{TomaJPB2002}.  

All $3p$-atomic delays computed using two-photon matrix elements with RPAE correlation, 
labelled as ($*,+,\times$) in Fig.~\ref{fig_delay}~(a), 
show almost identical atomic delay curves.  
It is, therefore, plausible that correlation with inner orbitals are not critical for the atomic delay from the $3p$ orbital, 
except at: 
(i) the Cooper minimum ($\sim 52$\,eV) and 
(ii) the opening of the Ar$3s^{-1}$ channel ($\sim 28$\,eV). 
The atomic delays from the Muller potential are in
qualitative agreement with the correlated results both
showing negative delay peaks (so-called `advances') at roughly the position of the Cooper minimum shown in Fig.~\ref{fig_cross}(a).  
Quantitatively, the delays from the Muller potential differ by up to $50\,$as compared to the correlated models.
The atomic delays calculated using only `forward'-type Coulomb interactions  
(magenta $\opendiamond$),
i.e. restricted to the perturbation diagrams in Fig.~\ref{fig_fforwardbbackward}(a)--(d), % in Fig.~\ref{fig_delay}~(a), 
show that also ground-state correlation from RPAE is important to correctly position the delay peak and to reach an accuracy beyond a few tens of attoseconds. 
All models predict that the delay peaks in Fig.~\ref{fig_delay}~(a) are {\it negative}, which implies that the photoelectron escape {\it faster} due to the Cooper minimum as a result of a $-\pi$-shift of the dipole phase. 
The Cooper minimum is, in this sense, behaving in the opposite way as compared to 
an autoionizing resonance that holds the photoelectron close to the atom for an extended time, leading to a delay of emission, as shown by Eq.~(\ref{resdelay}). 

The $3s$-atomic delays in photoionization,
shown in the Fig.~\ref{fig_delay}(b),
are quite different from those from the $3p$-orbital. 
In particular, the minimum in the $3s$--partial cross-section in Fig.~\ref{fig_cross}~(b) are replaced by a large positive delay peaks for all RPAE models including inter-shell correlation (black $*$, blue $+$ and magenta $\opendiamond$). 
It has been pointed out earlier that inter-shell correlation is required to produce such a delay peak \cite{GuenotPRA2012, KheifetsPRA2013}. 
In agreement with this statement, our result restricted to $3s$ intra-shell correlation (red $\times$% in Fig.~\ref{fig_delay}~(b)
) shows an atomic delay curve without a peak. 
Interestingly, the shape of the correlated peak changes quite significantly when the 
inter-shell correlation includes 
all atomic orbitals from $M$ and $L$-shells (black $*$),
as oppose to only the $M$-shell (blue $+$). This result implies that the atomic delay is sensitive also to inner-shell electrons. 
We have verified that including the correlation with $K$-shell electrons does not further modify the atomic delay.  
The fact that the correlated peaks are {\it positive} from $3s$ suggests that the photoelectrons are {\it delayed} by the correlation-induced Cooper minimum
and that the dipole phase increases by a $+\pi$-shift, in direct contrast to the case of the $3p$-orbital.  
At first glance such discrepancy is surprising if we consider the minimum from $3s$ orbital as a shifted replica of the Cooper minimum from $3p$ orbital. 

In our earlier work \cite{DahlstromPRA2012} we used the Hartree-Fock (HF) values for the binding potentials. The relatively large error of the HF $3s-$orbital energy, 
$\epsilon_{3s}^\mathrm{EXP}-\epsilon_{3s}^\mathrm{HF}$=5.5\,eV, can be corrected for empirically by adjusting the eigenvalues of the bound orbitals to the experimental energies \cite{Amusia1990}. 
In this work we use the experimental values 
$\epsilon_{3p}=-15.8$\,eV and 
$\epsilon_{3s}=-29.25$\,eV. 
As a consequence the data presented in Fig.~\ref{fig_delay} show some deviation with the results in Fig.~4~(b) from Ref.~\cite{DahlstromPRA2012}. 
In both cases, however, 
the delay peak is negative for the $3p-$orbital  
and positive for the $3s-$orbital. 

The different signs of the delay peaks in argon have been 
supported by RPAE calculations of the one-photon Wigner delay \cite{GuenotPRA2012,KheifetsPRA2013}. 
In contrast, the time-dependent local density approximation (TDLDA) shows that the delay peak from the $3s$-orbital should be negative (i.e. the dipole phase makes a $-\pi$-shift) \cite{DixitPRL2013}. The TDLDA calculation shows better agreement with the experimental measurements \cite{KlunderPRL2011,GuenotPRA2012} as compared to the RPAE results and a detailed survey of the two theoretical model would be desirable to understand this discrepancy. 
To this end, we have calculated the atomic delay using additional models and 
found that it is, indeed, possible to change the sign of the delay peak depending on the degree of correlation.   
In all cases we use the HF basis 
for the occupied orbitals and the HF plus projected potential for the excited states, 
see Eq.~(\ref{h_hf}) and (\ref{u_proj}), where the monopole interaction 
is chosen to be that of the final static hole, $v=3s$. 
Only correlation among the M-shell is included. 
A change of sign of the $3s$-delay peak is achieved when the correlation effects are restricted to include  
(i)  
the Tamm-Dancoff approximation (TDA) 
by solving the equations given by diagrams in 
Fig.~\ref{fig_fforwardbbackward}~(a), (b) and (d),
which corresponds to an infinite series of forward propagating bubble diagrams; and 
(ii) 
the random phase approximation (RPA) 
given by the diagrams in 
Fig.~\ref{fig_fforwardbbackward}~(a), (b), (d) and (f), 
which includes additionally diagrams for ground-state correlation from reverse propagating bubble diagrams. 
As expected, the delay peak from the TDA is shifted to higher photon energies due to the lack of ground-state correlation, while the RPA peak is located at a better position, 
see Fig.~\ref{fig_delay}(c).   
No delay peak from the $3s$-orbital is found when correlation is not included  
[Fig.~\ref{fig_fforwardbbackward}~(b)] 
or when only direct Coulomb interactions (so-called `ladder' interactions) are included
[Fig.~\ref{fig_fforwardbbackward}~(a), (b), (c)]. 
This is because the direct Coulomb interactions 
are too weak to induce the Cooper minimum from the $3p$-orbital 
over the direct ionization path from the $3s$-orbital. 
None-the-less, our result shows that the inclusion of direct Coulomb interactions in RPA, which is equivalent to the total RPAE approximation, are strong enough to change the sign of the $3s$-delay peak. 
We conclude that the $3s$-atomic delay is sensitive to correlation effects, such as the relative strength of direct and exchange interactions with the $3p$-orbital. 

In contrast, we have verified that neither TDA nor RPA change the sign of the delay peak from the $3p$ orbital in argon. This is because the $3p$-atomic delay peak comes mainly from the interference between the intermediate $s$ and $d$-waves that both stem from the $3p$ orbital. However, when we artificially compute the delay on the pure $d$-wave we find that the sign of the phase  peak changes from a negative $\pi$-shift (RPAE) to a positive $\pi$-shift (TDA and RPA). In this case, it is the correlation contribution from the $3s$-orbital that dominates close to the d-wave Cooper minimum and determines the sign of the $\pi$-shift. If correlation effects are excluded, the $d$-wave suffers a sudden $\pi$-shift and no preferred direction of the delay peak should be expected. The TDA and RPA results agree qualitatively with the TDLDA method that also predicts a positive $\pi$-shift for the $d$-wave, as seen in Fig.~3 in Ref.~\cite{DixitPRL2013}. We identify that the different trends obtained through the RPAE or TDLDA methods can be explained by small changes in the single electron-hole pair correlation model. It remains an open question if additional double-excited diagrams will cause the RPAE result to become more like RPA result.     

In Table~\ref{tab_res} we present numerical results for the atomic delays at a selection of RABITT sidebands where experiments have been performed \cite{KlunderPRL2011,GuenotPRA2012}.
\begin{table}[h]
\caption{\label{tab_res}Atomic delay, $\tau_\mathrm{A}$, for argon from $3p$ and $3s$-orbitals, where the curly bracket indicates the correlated orbitals included in the RPAE and RPA model.}
\begin{indented}
\item[]\begin{tabular}{@{}llll}
\br
$\tau_\mathrm{A} (as)$ & SB:22 & SB:24 & SB:26 \\
\mr
RPAE\,3p\{all\}         &   6 & -2  & -14 \\
RPAE\,3s\{M-shell\}     & -43 &  23 & 329 \\
RPA\,3s\{M-shell\}      & -75 & -67 & -87 \\
\br
\end{tabular}
\end{indented}
\end{table}
The $3p$-atomic delays are small compared to the $3s$-atomic delays in this energy region, which implies that the difference in atomic delay between the two orbitals is dominanted by the $3s$-delay. At sideband 22, the difference between the $3s$ and $3p$-atomic delay has been determined to be $-80\pm50$\,as \cite{GuenotPRA2012}. The value of the RPAE approximation is $-49$\,as, while the RPA yields $-81$\,as. At sideband 24, the difference in delay is $-100\pm50$\,as \cite{GuenotPRA2012}, which is far from the RPAE result ($25$\,as) but closer to the RPA result ($-65$\,as). Unfortunately, the experimental data at sidebands 26 is confirmed by neither RPAE nor RPA. We believe that the better performance of the RPA model is accidental and that it is the neglected double excited states that are causing the discrepacy, as argued by Carette et al. \cite{CarettePRA2013}. Including double-excited states using diagrammatic perturbation theory is beyond the scope of this article, but it will be the topic of future works. 

\subsection{Dependence on detection angle for atomic delay measurements}
\label{sec_detang}

\begin{figure}[h]
	\centering
	\includegraphics[width=\textwidth]{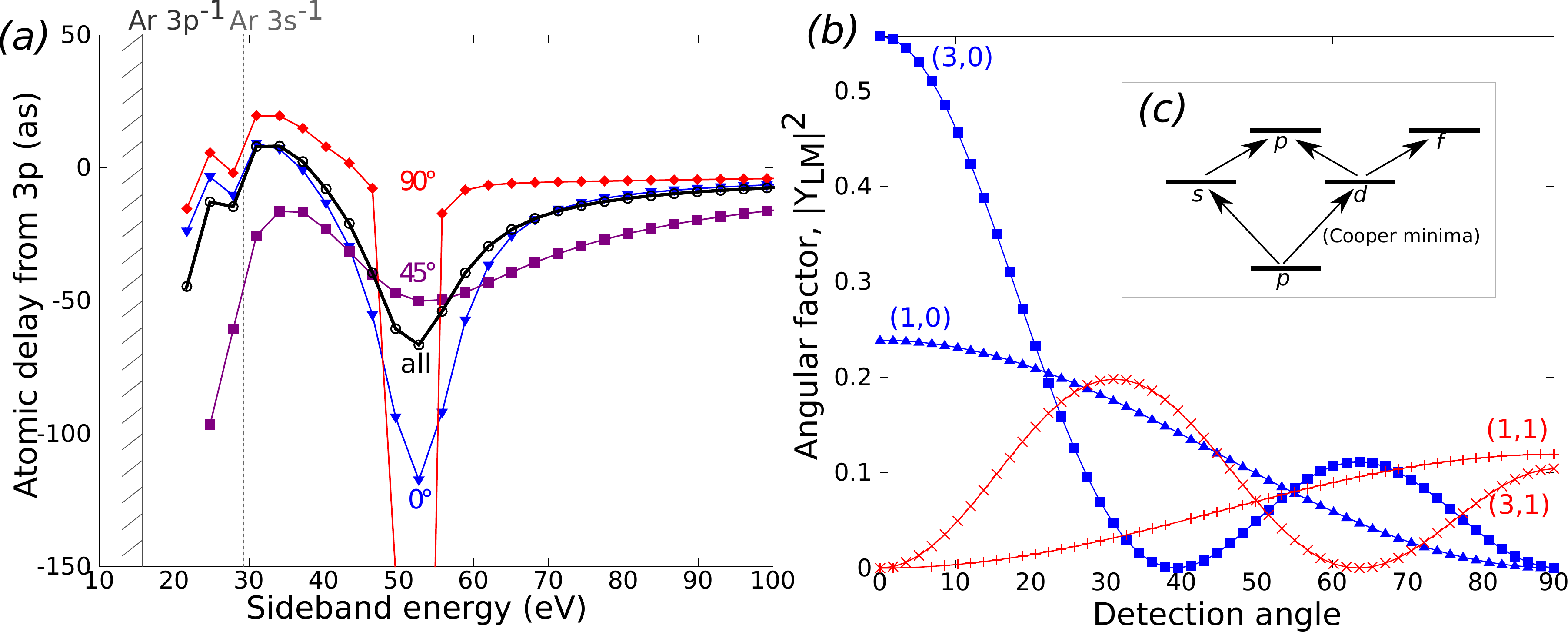}
	\caption{\label{fig_ang} 
(Color online) 
(a) Atomic delay from Ar$3p^{-1}$ at photoelectron detection angles: 
$\theta=0^\circ,\,45^\circ$ and $90^\circ$ labelled by 
blue $\opentriangle$, purple $\opensquare$ and red $\opendiamond$, respectively. 
The atomic delays with angle-integrated detection [Eq.~(\ref{tauenergy})] is labelled by black $\opencircle$. The IR photon is 1.55\,eV.
(b) Squared spherical harmonics for the corresponding final partial wave states with notation $(L,M)$. (c) Angular momentum paths for two-photon process staring in a $p-$state.
} 	
\end{figure}

The choice of method for detecting the photoelectrons may strongly influence the observed atomic delay.    
In Fig.~\ref{fig_ang}~(a) we show how the delay from the $3p$-orbital in argon varies as a function of XUV photon energy for a range of different detection angles (inter-shell correlation among the $L$ and $M$-shells is included).
%
%
% Ar3p m=0,1 RPAE (EXP)
The atomic delays are calculated using Eq.~(\ref{taumomentum}) for angles $\theta=0^\circ,\,45^\circ$ and $90^\circ$. 
At $0^\circ$ (blue $\bigtriangledown$) the negative delay peak at the Cooper minimum is $-117\,$as. 
At $45^\circ$ (magenta $\opensquare$) the peak is flatter, with a value of $-50\,$as, 
while the delay close of the ionization threshold becomes steeper. 
At $90^\circ$ (red $\opendiamond$) the delay shows an exceptionally sharp spike 
that extends to $-467\,$as
% (or $868\,$as with the half period of the IR field RABITT modulus: $2.67/2=1.34\,$fs)
. 
%
%Using the Muller potential and the $M-$shell only places the peak slightly lower at $613$ and $737\,$as, respectively. 
%
%
The energy-integrated delay (black $\opencircle$) evaluated using Eq.~(\ref{tauenergy}), is quite close to the angle-resolved delay for $\theta=0^\circ$, except in the vicinity of the Cooper minimum. This implies that the one intermediate partial wave is dominant at photon energies away from the Cooper minimum \cite{DahlstromCP2013}. 

In more detail, the delay variation can be understood from the angular distributions of the photoelectron wave packet. 
The allowed angular momentum paths $(l_a\rightarrow l_p\rightarrow l_q)$ are:
$(p\rightarrow s\rightarrow p)$;
$(p\rightarrow d\rightarrow p)$; and
$(p\rightarrow d\rightarrow f)$, as shown in Fig.~\ref{fig_ang}~(c). 
The probability for emission is proportional to the squared spherical harmonic of the final angular momentum, as is evident from Eq.~(\ref{final}). 
In Fig.~\ref{fig_ang}~(b), we plot $|Y_{LM}(\mathbf{\hat k})|^2$ for all final partial wave states as a function of the angle, $\theta$, from the polarization axis of the fields.
For parallel detection ($\theta=0^\circ$) only the $M=m_a=0$ states are detected: $(L,M)=(1,0)$ and $(3,0)$, while for perpendicular detection ($\theta=90^\circ$) only $M=m_a=\pm 1$ states are detected: $(1,1)$ and $(3,1)$.  
Our results are in good qualitative agreement with the respective delays of specific final partial wave states calculated by Mauritsson et al. \cite{MauritssonPRA2005}.
%In the intermediate region ($\theta\approx 45^\circ$), the atomic delay will depend sensitively on the detection angle because all final states contribute to the emission. 
% 
When the intermediate $d$-wave passes through the minimum, 
located at $\hbar\Omega\approx53\,$eV [see Fig.~\ref{fig_cross}~(a)],  
only the path through the $s$-wave will remain open. 
In parallel detection it is possible for the electron to transition through an $s$-state, which smears out the $\pi$-shift and makes the delay peak not so steep. 
In contrast, the perpendicular detection of an electron requires $M=\pm1$, which makes it impossible to transition via the $s$-wave, so that the background signal is strongly suppressed and the $\pi$-shift becomes abrupt and the delay peak becomes more sharp. 
Assuming an abrupt $\pi$-shift from the Cooper minimum between the XUV harmonics, the delay can be estimated as $\tau_\mathrm{Cooper}=\pm\pi/(2\omega)=\pm 667\,$as, which is in reasonable agreement with the sharp delay spike. The only possible  correlation contribution in this case is due to the $2p$-orbital in the $L$-shell.  
The detection of such sharp peaks are difficult to measure experimentally due the vanishing photoionization signal close to the absolute Cooper minimum.

\subsection{Role of Fano resonances}
\label{sec_fano}
%
% TALK ABOUT: FANO RESONANCE (3s-Rydberg series -> 3p)

\begin{figure}[ht]
	\centering
	\includegraphics[width=\textwidth]{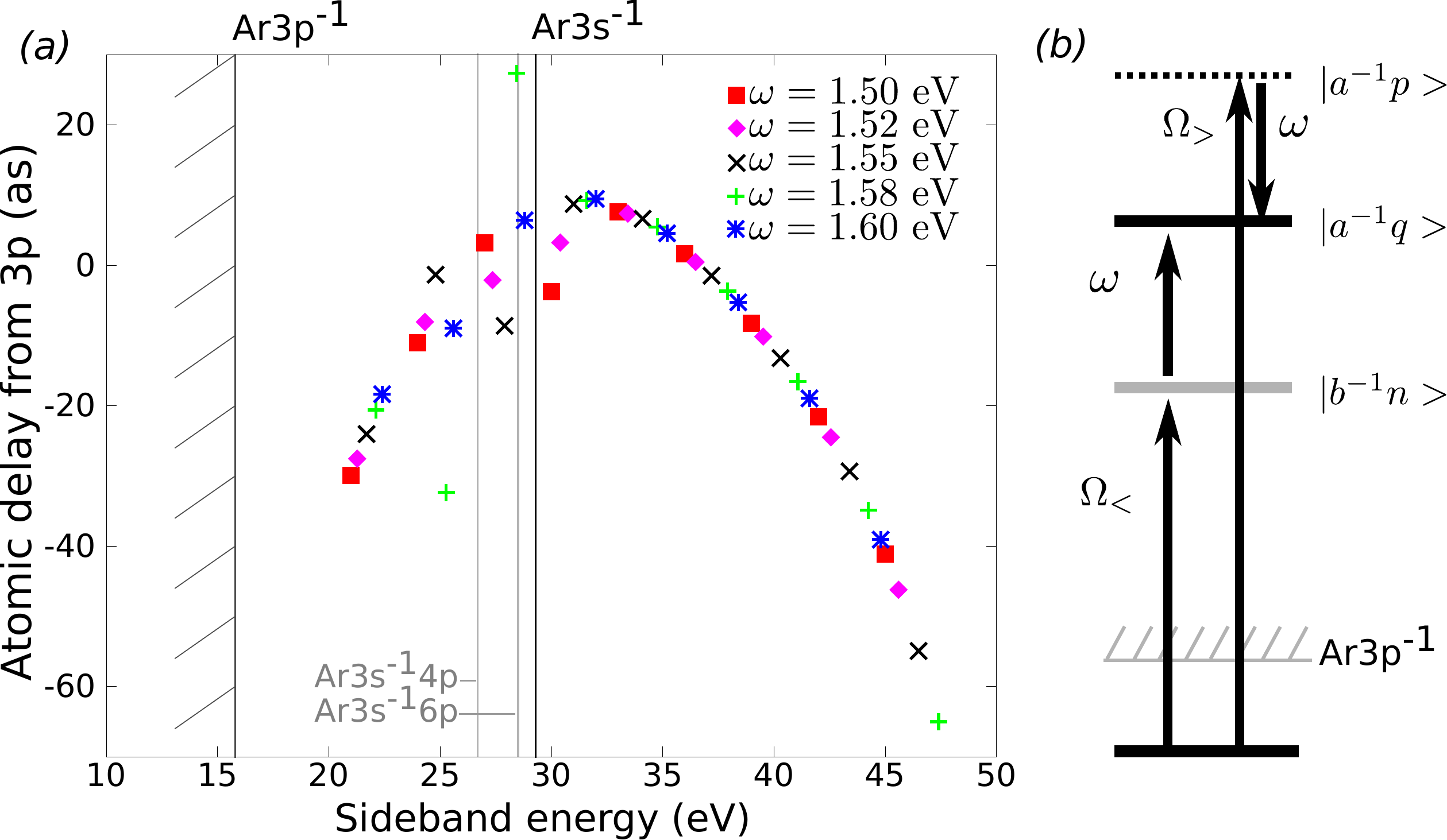}
	\caption{\label{fig_Ar_fano} 
(Color online) (a)
Atomic delay from Ar$3p^{-1}$ close to the Ar3s$^{-1}$ threshold for detection along the polarization axis. The strong delay variations are attributed to Fano resonances (Ar $3s^{-1} np$). (b) Schematic photon diagram where the absorption path, 
$\Omega_<+\omega$, 
is close to the resonance (grey bar), while the emission path $\Omega_>-\omega$, is far from the resonance.
} 	
\end{figure}

In connection with Fig.~\ref{fig_delay}~(a), we noted that the delay curves including inter-shell correlation (black $*$, blue $+$ and magenta $\opendiamond$) exhibit a peculiar feature, located on the 18th sideband ($\sim 28\,$eV above the atomic ground state).
The observed delay effects are due to the autoionizating Rydberg series leading to the opening of the Ar$\,3s^{-1}$ channel%, c.f. Ref.~\cite{SorensenPRA1994}
, see the schematic plot in Fig.~\ref{fig_Ar_fano}(b), 
where $a$ and $b$ are the $3p$ and $3s$-orbitals, respectively. 
This conclusion is supported by 
the fact that no such delay features are present in neither 
the intra-shell RPAE model (red $\times$) nor single-active electron approximation (green $\opencircle$) in Fig.~\ref{fig_delay}~(a). 
The delay effects are shown in more detail in Fig.~\ref{fig_Ar_fano}, where we finely detuned the odd XUV harmonic comb: 
$\Omega_{2N+1}=(2N+1)\omega$, 
by changing the IR photon energy, $\omega$. 
As a result, we obtain different final energies $2N\omega$ for the $2N$ sideband as a function of the IR photon energy. As mentioned above, we are interested here in the atomic effects and consider a monochromatic IR field for simplicity. A detailed comparison with experimental data should also take the convolution of the incident fields into account, as well as the energy resolution of the photoelectron spectrometer.  
First, we note that the delay curves align nicely in the regions far away from the autoionizing resonances. 
Close to the opening of Ar~$3s^{-1}$ channel, however, the atomic delay  exhibits strong variations depending on the detuning of the XUV and IR fields. 
The resonances also affect the atomic delay above the threshold of the Ar~$3s^{-1}$ ion, when the IR frequency is tuned to $1.5\,$eV.	 
The results presented in Fig.~\ref{fig_Ar_fano} are of qualitative character because it has been shown that the RPAE model does not produce quantitative Fano parameters of this autoionizing Rydberg series \cite{Amusia1981,Amusia1982}. The simple analysis for RABITT measurements in Sec.~\ref{sec_res} seems to hold up fairly well. The adjacent sidebands 16 and 18 with $\omega=1.58$\,eV exhibit large negative and positive delay, respectively, which implies that harmonic 17 must be close a resonance [see green $+$ close to Ar3s$^{-1}$4p in Fig.~\ref{fig_Ar_fano}(a)]. Similar effects are also observed at other resonances, e.g. for sideband 18 and 20 with $\omega=1.5$\,eV where harmonic 19 is close to Ar3s$^{-1}$6p.   
The fact that the atomic delay is so sensitive to the detuning of the XUV harmonics 
indicates a breakdown of Eq.~(\ref{delays}). As mentioned in Sec.~\ref{sec_res}, 
it is the frequency of the IR photon that is too large to resolve the narrow  
spectral phase variation of the autoionizing resonances, 
c.f. Eq.~(\ref{fano_phase}). 
This implies that the first term in Eq.~(\ref{delays}),
that used to be the Wigner delay $\tau_\mathrm{W}$, 
is no longer related to the delay of the photoelectron motion. 
This is because the {\it finite difference approximation} to the Wigner delay, 
$\Delta\eta/\Delta\omega$ with energy difference $\Delta\omega=2\omega$, 
is not small enough to resolve the spectral variation around the resonances.
A remedy for this might be to use probe fields with smaller photon energy. 
A first step in this direction is given in the next subsection. 

\subsection{Dependence on probe-field wavelength}
\label{sec_probewave}
% Ne2p w/ 800 nm , 1.3 mu and 2 mu RPAE (EXP)
% Ar3p w/ 800 nm , 1.3 mu and 2 mu RPAE (EXP)

\begin{figure}[ht]
	\centering
	\includegraphics[width=\textwidth]{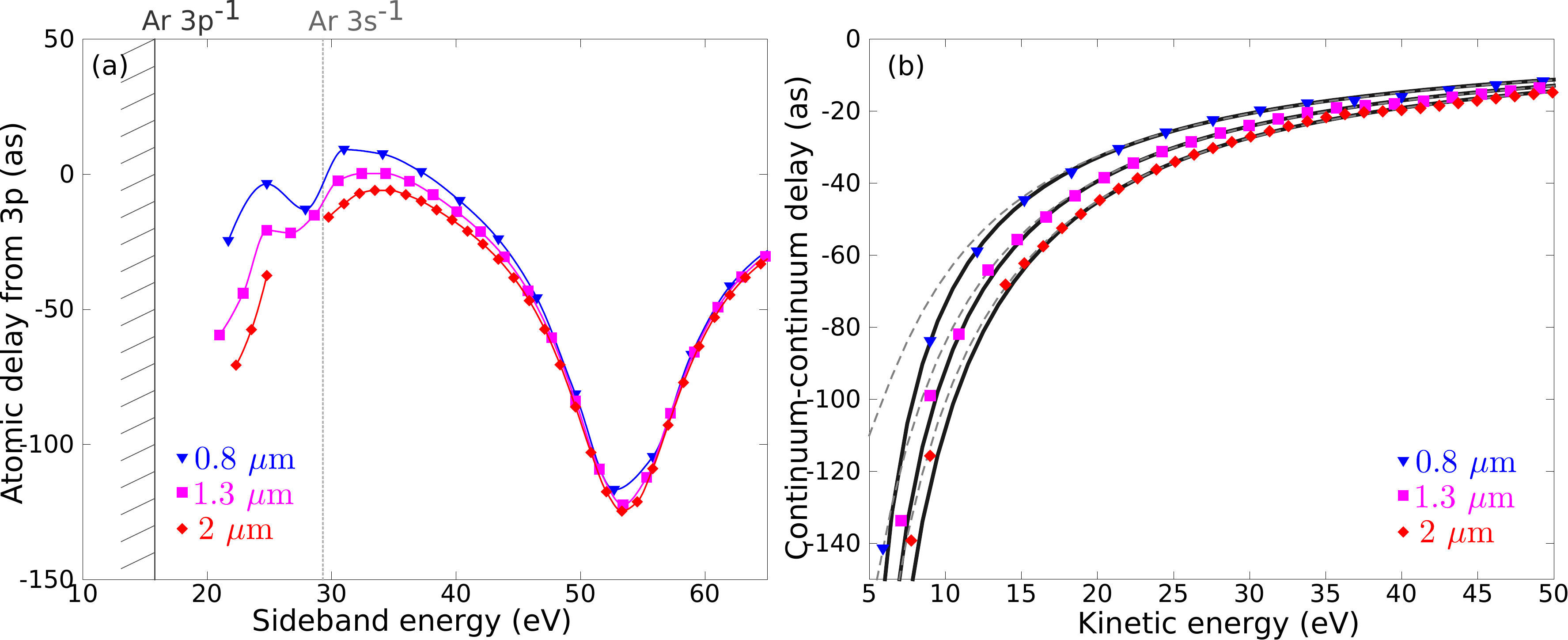}
	\caption{\label{fig_Ar_wav} 
(Color online)
Atomic delay, $\tau_A$, from Ar$3p^{-1}$ observed along the polarization axis using three different IR laser photons:
blue $\bigtriangledown$, 
magenta $\opensquare$ and red $\opendiamond$ 
corresponding to 0.8, 1.3 and 2\,$\mu$m wavelength, 
respectively. 
(b) Continuum--continuum (CC) delays are extracted using: 
$\tau_\mathrm{CC}=\tau_\mathrm{A}-\tau_\mathrm{W}$ 
and show good agreement with the asymptotic approximation 
from Ref.~\cite{DahlstromJPB2012}, shown as black full and grey dash curves.  
} 	
\end{figure}

In this subsection we present the atomic delay 
for a few different values of IR laser wavelengths, 
shown in Fig.~\ref{fig_Ar_wav}~(a). 
In order to resolve the Wigner delay of the autoionizing resonances 
the XUV harmonic comb spacing must typically be on the order of $10$\,meV or less. 
Here, we present results from a more modest decrease of the XUV harmonic comb spacing for 
the atomic delay from the $3p^{-1}$-orbital.   
A similar calculation has already 
been performed for hydrogen \cite{DahlstromCP2013}.
It is observed that the atomic delay, 
namely $\tau_\mathrm{A}$ from Eq.~(\ref{delays}), 
grows to more negative values as the wavelength is increased. 
The change in $\tau_\mathrm{A}$ is most substantial for a
low kinetic energy of the photoelectron.   
In Fig~\ref{fig_Ar_wav}~(b) we show that 
$\tau_\mathrm{cc}$ from Eq.~(\ref{delays}) 
is a negative monotonic function 
that approaches zero as the kinetic energy of the photoelectron is increased, 
in agreement the earlier work on hydrogen 
\cite{DahlstromCP2013} and noble gas atoms \cite{DahlstromPRA2012}. 
We have verified that the asymptotic approximation for the CC delay (see Eq.~(100) in Ref.~\cite{DahlstromJPB2012}) is in excellent agreement with the numerical data at high kinetic energies, see dashed curves in Fig.~\ref{fig_Ar_wav}~(b). 
Interestingly, the CLC delay surmised by Pazourek et al. in Ref.~\cite{PazourekFD2013} is indistinguishable from this CC approximation (i.e. the dashed curves).
At low kinetic energies, the numerical results deviate from the simple analytical formulas and more accurate analytical models are required. The use of incomplete gamma functions, see solid black curves in Fig.~\ref{fig_Ar_wav}~(b), provides an improved estimate of $\tau_\mathrm{cc}$, but further matching to numerical results might be needed for even smaller probe photons.  
In the pursuit of atomic delay measurements using narrow spaced XUV harmonics and extremely low energy probe photons, we must expect increasing negative contributions from the CC-delay. 
% 

% OUTLOOK
%\input{corr_beyond_rpae.tex}

% CONCLUSION
\section{Conclusion}
\label{sec_conc}

In this paper we have given a detailed account of how two-photon two-color above-threshold  matrix elements can be used to compute the atomic delay in photoionization from noble gas atoms. 
The numerical results were focused on the RABITT method 
\cite{MullerAPB2002,TomaJPB2002}, 
but it is expected that the angle-resolved atomic delays are 
directly relevant also for accurate calibration of 
the (I)PROOF method
\cite{ChiniOE2010,LaurentOE2013}
and the attosecond streak-camera method  
\cite{ItataniPRL2002,MairessePRA2005}. 
A systematic analysis of laser-assisted photoionization from the $M$-shell of argon atoms has been presented with the aim to disentangle different correlation contributions. In this work, many-body screening effects were included to the RPAE level on the XUV photon. The so-called ``continuum--continuum delay'' induced by the IR field was found to be in good agreement with earlier calculation for hydrogen \cite{DahlstromCP2013}. It was found that the correlation effects were dominated by intra-shell interactions in photoionization from the outer $3p$-orbital. In contrast, inter-shell interactions had to be included in order to describe the photoionization from the inner $3s$-orbital. The most important screening effect came from the outer $3p$-orbital, which was well-known from earlier work on one-photon ionization \cite{AmusiaPLA1972}. Close to the $3s$-photoionization minimum, the atomic delay showed a large {\it positive} peak, in agreement with earlier work based on RPAE \cite{GuenotPRA2012,DahlstromPRA2012,KheifetsPRA2013}. Here, we demonstrated that also inner electrons, from the $L$-shell, show some sizable contribution to the delay peak.  
In comparison with experimental atomic delay data \cite{KlunderPRL2011,GuenotPRA2012} our results did not agree as well as the TDLDA method \cite{DixitPRL2013}, which, instead, predicts a {\it negative} delay peak from the $3s$-orbital. 
This appearent failure of the RPAE model was surprising because the performance of both RPAE \cite{KheifetsPRA2013} and TDLDA \cite{DixitPRL2013} is good when compared to experimental cross-section data \cite{MobusPRA1993}. 
Here, we identified that the $3s$-atomic delay made a dramatic change from a positive peak to a negative peak when the RPAE model was replaced by the more basic RPA model (or Tamm-Dancoff approximation, TDA). This demonstrated that the atomic delay measurements can be sensitive to small changes in the correlated interactions. A detailed comparative study of the models would be desirable. The use of diagrammatic methods has opened up for a unique way to identify the role of various classes for correlated interactions. 
The sign of the delay peak is related to the direction of the $\pi$-shift of the one-photon XUV matrix element, which is a topic long-standing interest because it can not be determined using conventional angle-resolved photoelectron emission. 
A similar study of the delay peak from the $3p$-orbital showed that RPAE, RPA and TDA all show negative delay peaks, independent of the level of correlation included. This was expected since the delay in this case is caused by an interference effect between the partial $s$ and $d$ waves from the initial $3p$-orbital (and not by detailed correlation effects). In fact, we also showed that the negative delay peak was reproduced using a single-active electron potential \cite{MullerPRA1999}.   

Further, a simple interpretation of the atomic delays observed close to  autoionizing resonances was presented. The implications of this model was that the usual interpretation of atomic delay measurements breaks down when the IR photon energy is comparable or larger than the energy width of the atomic resonances. Qualitative numerical results for the Ar$3s^{-1}np$ autoionizing Rydberg series were used to confirm our prediction. 
Similarly, the atomic delay detected by photoelectrons with angle-resolved  momentum was found to differ from the angle-integrated case. However, in the particular case where one intermediate (or final) partial wave was dominant the angle-resolved delay along the polarization axis was in good agreement with the angle-integrated delay. 

Although this work was limited to correlation effects based on the RPAE model, we stress that double excitations leads to additional autoionizing resonances located at higher photon energies. Such effects can be included in the diagrammatic framework and will be the topic of forthcoming work. In general, multi-configuration effects are expected to be important in the photoionization from the $3s$-shell due to a large coupling to double excited states. The multi-configurational approach for atomic delays \cite{CarettePRA2013} has been used to produce good agreement with the experimental atomic delay data, which also encourages further development of this computer program.  
%

% ACKNOWLEDGEMENT
\section*{Acknowledgement:}
Financial support from 
the Swedish Research Council (VR)
is acknowledged. 
%

% REFERENCES

%\bibliography{references}

\bibliographystyle{unsrt}

\end{document}